\documentclass[conference]{IEEEtran}
\usepackage[utf8]{inputenc}
\usepackage{url}
\usepackage{needspace}
\usepackage{mathtools}

\usepackage{algorithmic}
\usepackage{algorithm}
\usepackage[justification=centering]{caption}
\usepackage{subcaption}
\usepackage{placeins}

\usepackage{subcaption}
\usepackage{xcolor}
\usepackage{makecell}
\newcommand{\vew}[1]{
	\edef\start{\the\pgfmatrixcurrentrow-\the\pgfmatrixcurrentcolumn}
	\edef\end{\the\numexpr#1+\pgfmatrixcurrentrow\relax-\the\pgfmatrixcurrentcolumn}
	\expandafter\expandafter\expandafter\vewexplicit\expandafter\expandafter\expandafter{\expandafter\start\expandafter}\expandafter{\end}
}
\newcommand{\vewexplicit}[2]{
	\arrow[from=#1,to=#2,arrows,decorate,decoration={snake,amplitude=1pt,segment length=6.5pt}] {}
}

\usepackage{multirow}
\usepackage{amssymb}% http://ctan.org/pkg/amssymb
\usepackage{pifont}% http://ctan.org/pkg/pifont
\usepackage{siunitx}

\newcommand{\Sum}[2]{\displaystyle\sum_{#1} ^{#2}}

\newcommand{\rx}{$R_X(\theta)$}

\newcommand{\ry}{$R_Y(\theta)$}

\newcommand{\rz}{$R_Z(\theta)$}

\begin{document}
\renewcommand{\thefootnote}{\fnsymbol{footnote}}
% \bstctlcite{IEEEexample:BSTcontrol}

% \title{Local and global controller microcode \& QISA for Diamond NV center based algorithms}
% \title{ISA simulator for diamond NV center based quantum computers}
%\title {Microarchitecture Simulator for Fault-Tolerant Designs in Distributed Error-Detection Surface Code with Color Center}
% \title{Hardware-Software co-design for parallelising arbitrary quantum algorithms execution on distributed systems}
% \title{Hardware-Software Co-Design for Parallel Execution of Arbitrary Quantum Algorithms on Distributed Systems}
\title{Parallelizing Program Execution on Distributed Quantum Systems via Compiler/Hardware Co-Design}
% \author{Folkert de Ronde}
% \authornote{These authors contributed equally to this work.}
% \email{f.w.m.deronde@tudelft.nl}
% \orcid{0009-0003-2638-2916}
% \affiliation{
%     \institution{Delft University of Technology}
%     \country{The Netherlands}
% }

% \author{Alexander Knapen}
% \authornotemark[1]
% \email{aknapen@umich.edu}
% \orcid{0009-0009-5993-4899}
% \affiliation{
%     \institution{University of Michigan}
%     \country{USA}
% }

% \author{Stephan Wong}
% \email{j.s.s.m.wong@tudelft.nl}
% \orcid{0000-0003-3521-2612}
% \affiliation{
%     \institution{Delft University of Technology}
%     \country{The Netherlands}
% }

% \author{Sebastian Feld}
% \email{s.feld@tudelft.nl}
% \orcid{0000-0003-2782-1469}
% \affiliation{
%     \institution{Delft University of Technology}
%     \country{The Netherlands}
% }
% \renewcommand{\shortauthors}{De Ronde, Knapen, et al.}

\author{
    \IEEEauthorblockN{Folkert de Ronde\IEEEauthorrefmark{1}}
    \IEEEauthorblockA{
        Quantum \& Computer Engineering \\
        Delft University of Technology \\
        Delft, The Netherlands \\
        f.w.m.deronde@tudelft.nl
    }
\and
    \IEEEauthorblockN{Alexander Knapen\IEEEauthorrefmark{1}}
    \IEEEauthorblockA{
        Computer Science \& Engineering \\
        University of Michigan \\
        Ann Arbor, MI, USA \\
        aknapen@umich.edu
    }
\and 
    \IEEEauthorblockN{Stephan Wong}
    \IEEEauthorblockA{
        Quantum \& Computer Engineering \\
        Delft University of Technology \\
        Delft, The Netherlands \\
        j.s.s.m.wong@tudelft.nl
    }
\and

    \IEEEauthorblockN{Sebastian Feld}
    \IEEEauthorblockA{
        Quantum \& Computer Engineering \\
        Delft University of Technology \\
        Delft, The Netherlands \\
        s.feld@tudelft.nl
    }

    % show the equal-contribution message as a separate affiliation line (inside author{})
    \vspace{0.5em}
    \IEEEauthorblockA{\centering\small\IEEEauthorrefmark{1}These authors contributed equally to this work.}
}

% \and
% \IEEEauthorblockN{David Elkouss}
% \IEEEauthorblockA{Networked Quantum Devices Unit, Okinawa Institute of Science and Technology Graduate University, Okinawa, Japan\\
% QuTech, Delft University of Technology, Delft, The Netherlands \\
% david.elkouss@oist.jp}}
%\and
%\IEEEauthorblockN{James Kirk\\ and Montgomery Scott}
%\IEEEauthorblockA{Starfleet Academy\\
%San Francisco, California 96678--2391\\
%Telephone: (800) 555--1212\\
%Fax: (888) 555--1212}}

\maketitle
% \newpage
\thispagestyle{plain}
\begin{abstract}

% Quantum computers are rapidly developing. Therefore control hardware and compilers are needed that can support the development. 
As quantum computers continue to improve and support larger, more complex computations, smart control hardware and compilers are needed to efficiently leverage the capabilities of these systems. 

This paper introduces a novel approach to enhance the execution of quantum algorithms on distributed quantum systems. The proposed method involves the development of a hardware design that supports parallel instruction execution and a compiler that modifies the order of instructions to increase parallelism opportunities. The hardware design can be flexibly configured to facilitate parallel execution of instructions that have identical parameters. Furthermore, the compiler uses the underlying hardware constraints to intelligently reorder and decompose instructions to avoid dependencies.

The compiler, hardware, and their combination are evaluated using a runtime calculator and a benchmark quantum algorithm set. The results demonstrate a significant speedup, achieving a maximum average speedup of 16.5x and a maximum single-benchmark speedup of 56.2x relative to a baseline, serial execution model. Furthermore, we show a speedup can be obtained across all benchmarks using any of the proposed hardware schemes, although the degree of speedup is largely dependent on the type of quantum algorithm. Taken together, the results of this paper represent a significant step towards realizing high-performance quantum computing systems.

\end{abstract}

\section{Introduction}

% In classical computing, execution times of algorithms have been tried to lower for the past decades. This has been done by executing instructions at the same time (parallelism) or starting the execution of an instruction before the previous instruction has finished (pipelining). 
% % parallelism and pipelining have long been used to decrease algorithm execution time. 
% This parallelism and pipelining can be achieved by distributing tasks across multiple cores or stages. These same principles can be used in order to reduce execution times of quantum algorithms.

% In classical computing, reduction in execution times has been achieved for a long time. This reduction can be achieved by execution instructions of a classical algorithm at the same time (parallelism) or by starting the execution of a following instruction before the previous instruction has finished (pipelining). These techniques are both part of instruction-level parallelism (ILP).

In classical computing, techniques for executing multiple instructions at the same time, collectively referred to as \textit{instruction-level parallelism (ILP)}, have long been used to significantly decrease program execution time \cite{rau1993instruction}. Two of the most prominent ILP techniques employed in today's classical computing systems are pipelining and superscalar design. Pipelining divides the execution of instructions into independent steps that each use different hardware resources. This allows for overlapped processing of multiple instructions by performing different execution steps on each instruction at the same time. Building on this, superscalar designs copy hardware resources over multiple pipelines, called execution lanes, and execute different instructions in each lane simultaneously. With any ILP technique, the ability to execute two instructions in parallel is constrained by resource dependencies, in which both instructions require use of the same hardware resource at the same time, and logical dependencies between instructions, e.g., where the result of one instruction is required by the other instruction. To combat these constraints, compilers have been developed which are capable of reordering, or rescheduling, instructions such that resource and logical dependencies are avoided where possible and more ILP can be extracted from programs. 
% When an instruction finishes execution in a pipeline stage, the resources of that stage are freed up to be consumed by the following instruction. Hence, although only a single instruction completes per processor cycle, multiple instructions are worked on at the same time.
It is possible to 
 % In the future, it will become possible to
extend these ILP techniques to quantum computing systems as well, especially as quantum computing begins to transition into a modular, multiprocessor era \cite{ibmDevMap}. In particular, \textit{distributed quantum computing systems} have the potential to benefit greatly from ILP support in control architectures and compilers. In a distributed quantum computing system, the logical qubits specified in quantum circuits or programs are mapped onto multiple physical qubits on the quantum chip. Groups of these physical qubits are then operated on independently by separate hardware components managed by a control architecture (analogous to different execution lanes in a superscalar design). Since the hardware components per group are separate, the control architecture can, in principle, coordinate operations across different groups to be performed in parallel. In practice, a quantum compiler is needed to indicate to the hardware when parallel execution of operations is possible. To do so, the quantum compiler must be capable of decomposing gates on the logical qubits in a quantum circuit (logical instructions) into operations on the physical qubits (physical instructions) and rescheduling the operations in a way that avoids resource and logical dependencies.

Despite this promising potential, ILP techniques have not been widely implemented in quantum computing systems. At the level of the control architecture, efforts to facilitate parallel execution of quantum operations are typically restricted to frequency-multiplexed control \cite{asaadIndependentExtensibleControl2016} and readout \cite{risteMicrowaveTechniquesQuantum2020, heinsooRapidHighfidelityMultiplexed2018,ryanHardwareDynamicQuantum2017}. However, the maximum achievable parallelism in frequency-multiplexed systems is low due to the limited bandwidth of the transmission line cabling used, crosstalk effects, and leakage of qubits into higher energy levels \cite{risteMicrowaveTechniquesQuantum2020,heinsooRapidHighfidelityMultiplexed2018,vandijkElectronicInterfaceQuantum2019}. At the same time, quantum compilers such as Qiskit \cite{qiskit}, Cirq \cite{cirq}, ProjectQ \cite{projectQ}, and OpenQL \cite{OpenQL} do not consider opportunities for performing quantum operations in parallel when scheduling, nor do they take into account parallelization when decomposing logical instructions into physical instructions. 

Consequently, the resulting sequential execution of instructions significantly extends the total run time of quantum programs. However, since the underlying hardware is capable of supporting some level of parallel execution, this creates a large gap between the theoretically available performance in quantum computing systems and the performance that is achieved in practice. Furthermore, this performance gap will only widen as progress is made towards larger-scale systems performing computations on greater numbers of qubits. 

Therefore, in this paper, we propose and demonstrate hardware and compiler designs which exploit instruction-level parallelism to reduce execution times of quantum programs. The hardware design is capable of performing quantum operations in parallel, while the compiler uses rescheduling to increase the number of instructions that can be pipelined or parallelized.

The main contributions of this paper are:

% In this paper, we demonstrate the benefits of instruction-level parallelism in quantum computing systems by presenting a hardware design capable of performing quantum operations in parallel together with a compiler that uses rescheduling to increase the number of instructions in a quantum program that can be pipelined or parallelized. The main contributions of this paper are threefold: 
\begin{enumerate}
\item \textbf{Hardware Design:} we introduce a hardware design specifically tailored to distributed quantum systems. This design enables parallel execution for specific sequences of quantum instructions, and it can be configured to support varying degrees of parallelism.
\item \textbf{Quantum Compiler:} we present a quantum compiler that is aware of parallelism constraints in the underlying hardware design. Given a quantum program, the compiler reschedules its logical instructions and intelligently decomposes them into physical instructions such that, where possible, parallelism constraints are avoided and pipelined/parallel execution can be increased. 
% parallel and pipelined  identifies groups of parallelizable instructions within a quantum program and reschedules them to be executed in parallel by the hardware. Additionally, the compiler decomposes logical instructions into physical instructions, aiming to maximize parallelism and pipelining.
% \item \textbf{Evaluation Tool for Optimal Hardware Configuration:} we present an evaluation tool for identifying the hardware configuration for a set of quantum algorithm with minimal runtime. 
\end{enumerate}

The remainder of this paper is structured as follows. In Section \ref{sec:Sytem_background}, we introduce the theoretical background we created and used to design and create the hardware and compiler.
% relevant terminology created by us and conceptual frameworks we used and expanded upon in the rest of the paper.
In Section \ref{Sec:Hardware_design}, the design of the hardware is proposed based on the theory in Section \ref{sec:Sytem_background}, including the configuration parameters that can be used in order to achieve parallel execution. In Section \ref{sec:Compiler_design}, the compiler design used to increase parallel execution opportunities is presented. This section explains how the theory presented in Section \ref{sec:Sytem_background} can be leveraged by a compiler. In Section \ref{sec:runtime-analysis}, we present a run time analysis of instruction sequences that can be used to simulate the run time of quantum programs. In Section \ref{Sec:results}, the numerical speedup results achieved by our compiler and hardware co-design are presented. In Section \ref{sec:conclusion}, we present a conclusion about our work. 

% \input{Chapters/related_work}

% \section{Overview}
\section{Concepts and Terminology}\label{sec:Sytem_background}
In this section, we introduce concepts and terminology  we created/expanded upon that are important for understanding the design sections. We begin with a general description of the layout of distributed quantum computing systems and then consider the impacts of this layout on the performance of the system. Thereafter, we introduce two possible methods for mapping logical qubits specified in quantum programs onto physical qubits in the system. These mappings are then used to describe decompositions of quantum programs into operations that can be performed by the quantum computing system.

\subsection{Distributed Quantum Computing System Layout} \label{sec:system-layout}
\begin{figure}
    \centering \includegraphics[width=0.8\columnwidth]{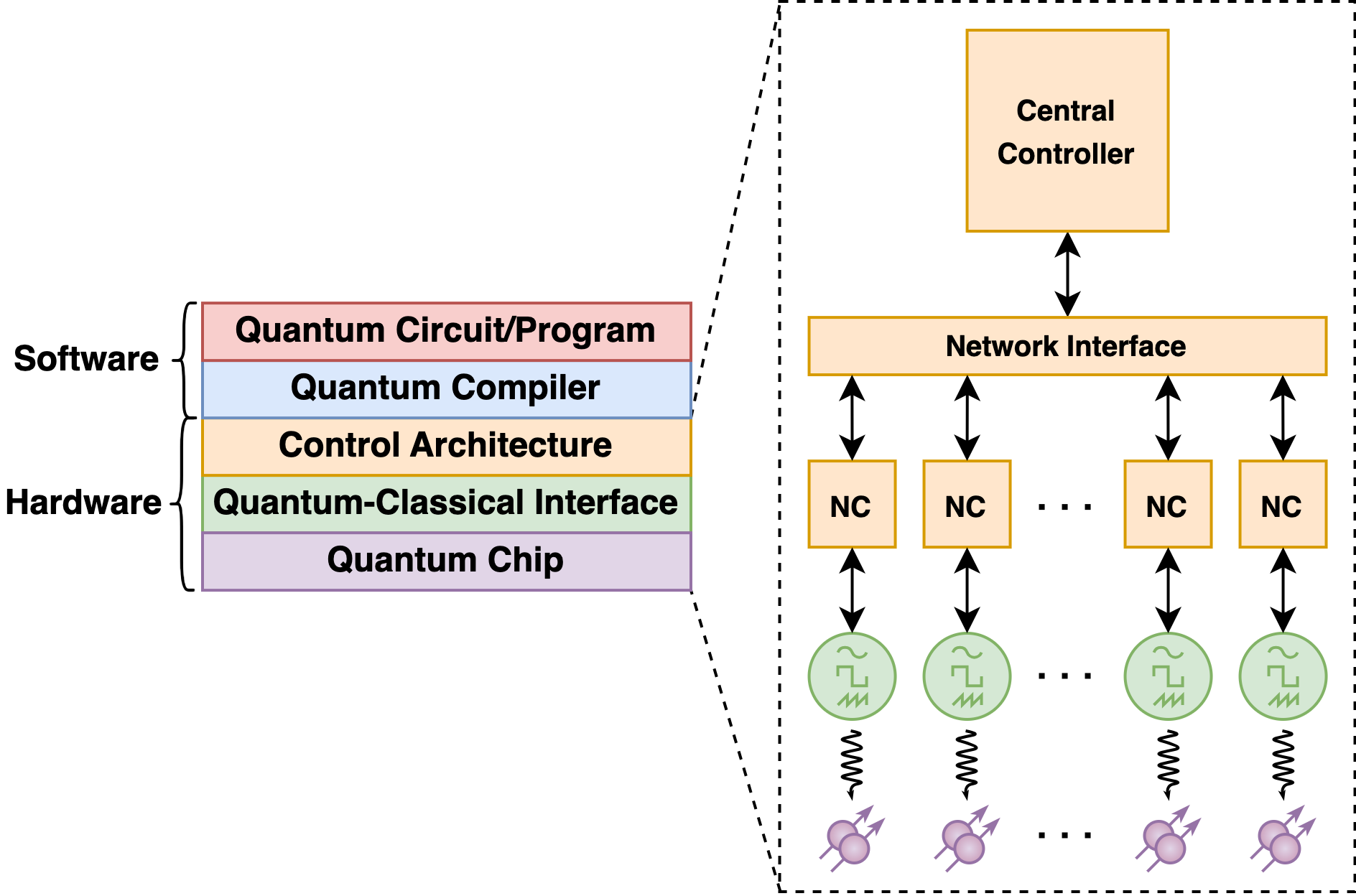}
    \caption{The quantum computing system stack (left-hand side) with a detailed structure of the control architecture, quantum-classical interface, and quantum chip layers (right-hand side).}
    \label{fig:control_structure}
    % \Description[]{A depiction of the different layers of the quantum computing stack from high-level software down to low-level hardware. The organization of the hardware layers for a distributed quantum computing systems is also shown.}
\end{figure}

The type of quantum computing system we consider in this paper is a distributed quantum computing system. Such a system can be described through stacked levels of abstraction (see Fig. \ref{fig:control_structure}) \cite{bandic2022full}. %\cite{fu2016heterogeneous}. 
% Each layer of the stack is explained next. 
At the top of the stack, quantum circuits, or programs, are used to represent quantum algorithms. These programs are specified at a purely logical level such that they are independent of the underlying qubit technology. For this reason, the gates and qubits in quantum programs are referred as \textit{logical gates} and \textit{logical qubits}, respectively. Then, the quantum compiler aims to improve the execution of the quantum circuit as well as decomposing the \textit{logical gates} into \textit{physical gates} supported by the underlying quantum technology. These top two layers of the stack consist only of software. However, the compiler bridges the gap between the software and hardware by further translating physical gates into instructions executable by the control hardware. 

The control hardware comprises the bottom three layers of the stack. First, a control architecture is used to orchestrate high-level program execution flow and to generate digital signals that specify how quantum operations should be performed. These operations are performed using the \textit{quantum-classical interface}, a collection of digital and analog hardware components responsible for qubit gate and readout signal generation. Finally, at the lowest level of the stack, the underlying \textit{physical qubit} technology (e.g., superconducting or diamond spin qubits) is housed in a quantum chip.

% One level down, a quantum compiler performs the translation of these ideal logical gates into gates that are physically realizable on the qubits in hardware. From this layer downward, gates and qubits are referred to as \textit{physical gates} and \textit{physical qubits}. The compiler generates the physical gates to perform, and which physical qubit(s) to perform them, on as instructions which are supplied to a hardware control architecture in the next layer of the stack. 

We now describe a particular organization of the control architecture and quantum-classical interface which is compatible with the system stack model; this organization is also depicted in Fig. \ref{fig:control_structure} on the right-hand side. The control architecture is organized as a central controller and a network of nodes which each contain a dedicated node controller (labeled NC). The central controller is used for high-level control of program execution flow, while the node controllers are used to configure and trigger the execution of specific quantum instructions by the quantum-classical interface (green disks in Fig. \ref{fig:control_structure}). The central controller utilizes a network interface to issue instructions to node controllers specifying which quantum operation is to be performed (the instruction opcode), how it should be performed (the instruction parameters, such as rotation angle), and which node controller(s) should perform it (the instruction address). Subsequently, if a node controller is addressed by an instruction, it is responsible for decoding the instruction into digital control values specifying how to synthesize quantum gates and perform qubit readout. Each node controller supplies these control values to its own set of signal generation hardware in the quantum-classical interface. Each set of signal generation hardware is in turn responsible for controlling a single group of physical qubits (purple spheres in Fig. \ref{fig:control_structure}).

\subsection{Address Encoding Methods} \label{sec:addressing-methods}
When evaluating the performance of a distributed quantum computing system, it is essential to consider the overhead in time required for transmitting each instruction's opcode, parameters, and address values over the network interface. This overhead is related to the cumulative width, or number of bits, used to encode an instruction's opcode, parameters, and address as well as the number of wires in the network interface used to transmit instruction data. Since each wire can transmit one bit per clock cycle, the overhead time required to transmit $W$ bits of the instruction over an interface of $L$ wires totals $\left \lceil \frac{W}{L} \right \rceil$ clock cycles. 

% The primary goal of the network interface design is to mitigate the performance penalty of the transmission overhead for each instruction. The most obvious mitigation technique is to directly minimize the transmission overhead by sending as few instruction bits as possible. On the other hand, we could instead issue multiple instructions at the same time to overlap the individual instruction overheads. The effectiveness of these techniques can be quantified through two performance metrics: \textit{single-instruction latency} (i.e., the amount of time required to complete a single instruction) and \textit{throughput} (i.e., the number of instructions being executed per unit time). 

The primary goal of the network interface design is to reduce the fraction of the total execution time spent on transmitting instructions from the central controller to the network of node controllers. The most obvious overhead reduction technique is to send as few instruction bits as possible, thus minimizing the \textit{single-instruction latency} (i.e., the amount of time required to transmit a single instruction). Moreover, multiple instructions could be issued over the network interface at the same time to overlap their individual transmission overheads in time. This corresponds to an increase in the \textit{throughput} of the system (i.e., the number of instructions being executed per unit time).

Both the throughput and single-instruction latency are highly dependent on how node controllers are addressed by the instructions issued from the central controller. The single-instruction latency can be reduced by utilizing fewer address bits in the instructions, while throughput can be increased by targeting multiple node controllers with a single instruction address. We now consider two common binary encoding methods to encode the instruction address information, one which minimizes the width of the address, and another which is capable of targeting multiple node controllers simultaneously.

\begin{figure*}
    \centering
    \begin{subfigure}{0.4\textwidth}
        \includegraphics[width=\textwidth]{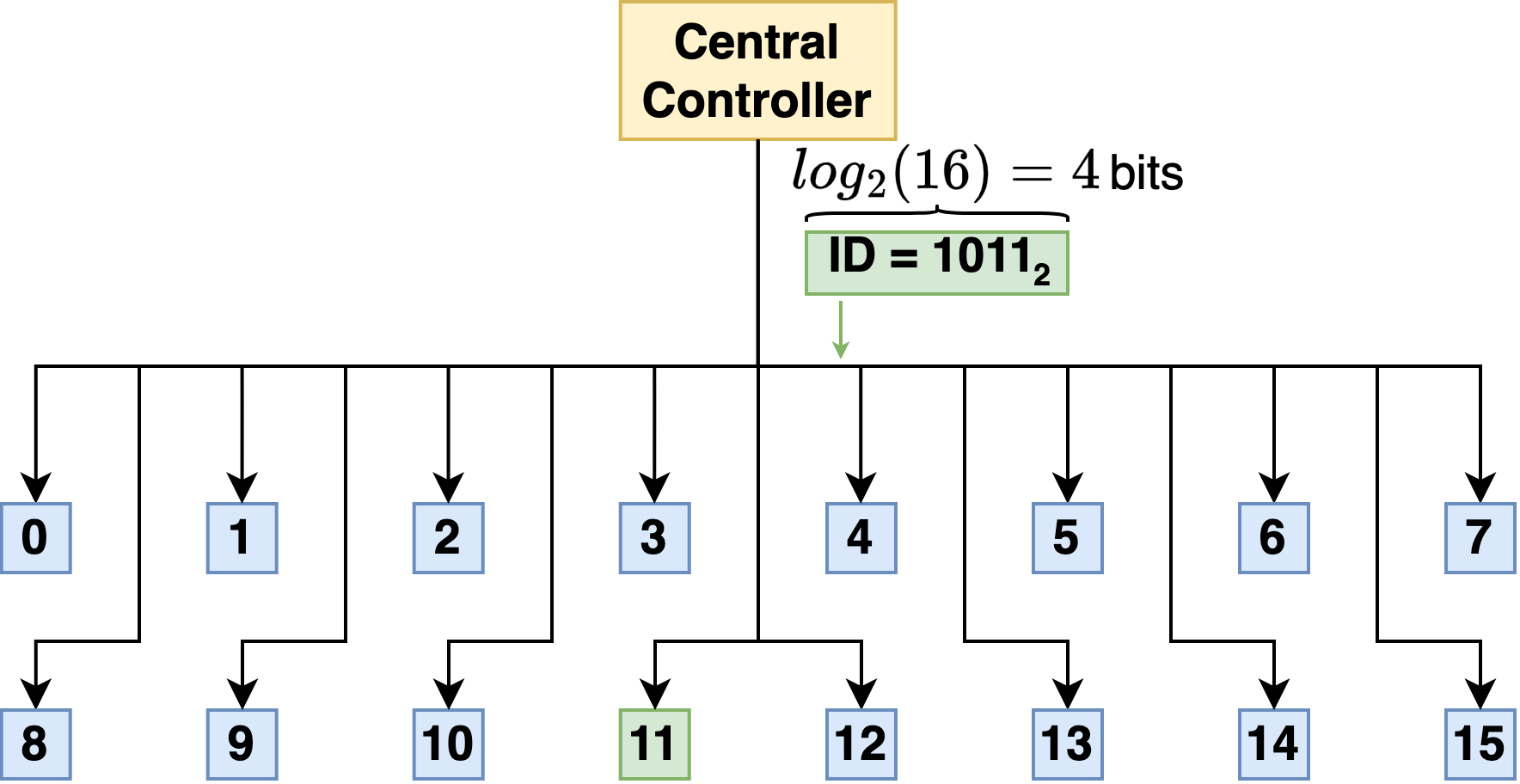}
        \caption{ID-encoded address.}
        \label{fig:id-address}
    \end{subfigure}\hspace{1cm}
    \begin{subfigure}{0.4\textwidth}
        \includegraphics[width=\textwidth]{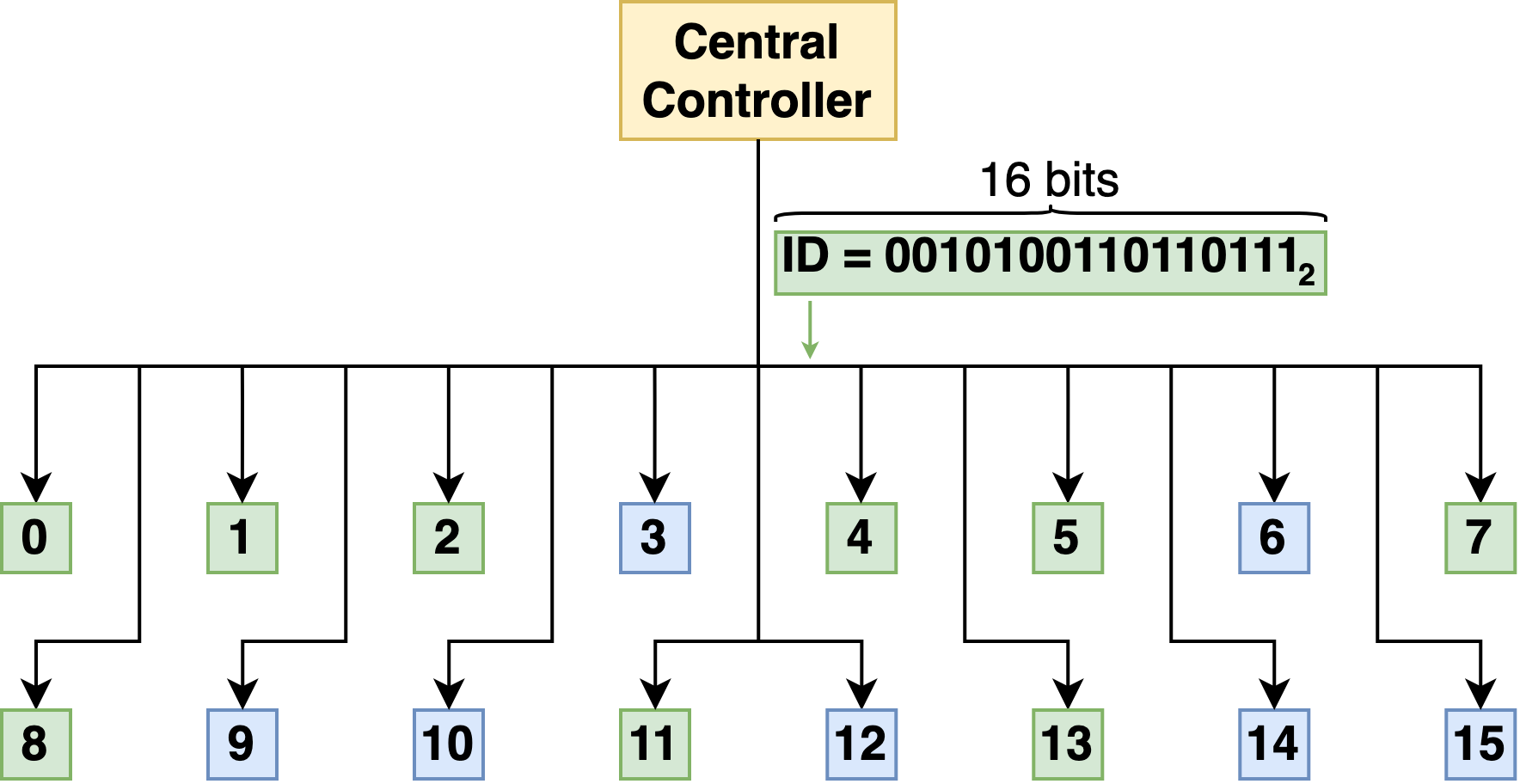}
        \caption{Bitmap-encoded address.}
        \label{fig:bit-address}
    \end{subfigure}
    \caption{Node controller selection using an ID-encoded address vs. a bitmap-encoded address. (a) An ID-encoded address has a smaller width but can only target a single node controller. (b) A bitmap-encoded address has a larger width but is capable of targeting multiple node controllers simultaneously.}
    % \Description[]{A diagram showing instruction address transmission over the network interface using an ID-encoded versus a bitmap-encoded address. The ID-encoded address requires fewer bits to be transmitted, but the bitmap-encoded address can target multiple node controllers simultaneously.}
    \label{fig:addr-methods}
\end{figure*}

The first and simplest addressing method encodes the address of an instruction as a unique ID referring to a single node controller (see Fig. \ref{fig:id-address}). For a network of $2^k$ node controllers, a $k$-bit address is needed to be capable of representing any node controller in the network uniquely. As a result, the address size, and thus the latency of transmitting the address over the network interface, grows logarithmically with the network size. While the small address size reduces the single-instruction latency, only one node controller is addressed per instruction. Therefore, the maximum throughput of the system is $1$, and algorithm execution is fully serialized. Such a system can be likened to a single-instruction, single-data (SISD) execution model in Flynn's taxonomy for classical computing systems \cite{flynn1966very}.

Alternatively, the address of an instruction can be encoded as a bitmap, where bit position $k$ of the address indicates whether or not the $k$th node controller in the network should execute the instruction (see Fig. \ref{fig:bit-address}). Using this addressing method, multiple node controllers can be issued instructions in parallel by setting their corresponding bits in the address bitmap. This is similar to a single-instruction, multiple-data (SIMD) parallel execution model in Flynn's taxonomy \cite{flynn1966very}, where the maximum throughput of the system grows with the size of the address bitmap. At the same time, when using a bitmap address encoding, the address size scales linearly with the number of node controllers in the network. Therefore, there will be an increase in the single-instruction latency associated with transmitting the larger address.

From this analysis, it is clear that a trade-off exists between the single-instruction latency and the maximum achievable throughput of the system. Namely, increasing the size of the address bitmap allows for more node controllers to be addressed simultaneously, but it also increases the single-instruction latency. We introduce two address-encoding-specific properties to capture these competing effects. The first property, $\rho$, specifies the maximum number of node controllers that can be addressed in a single instruction using a given address encoding method. For this reason, we refer to $\rho$ as the \textit{parallelizability factor}. The second property, $\delta$, denotes the number of additional clock cycles required to transmit an instruction using the given address encoding method relative to a baseline SISD execution model; $\delta$ is referred to as the \textit{parallelization overhead}. In the context of the quantum computing system's performance, an address encoding method with a larger $\rho$ value and lower $\delta$ value is preferred.

% \begin{figure}
%     \centering
%     \includegraphics[width=\columnwidth]{Figures/address_schemes.png}
%     \caption{Node controller selection using an ID-encoded address vs. a bitmap-encoded address.}
%     \label{fig:addr-methods}
% \end{figure}

\subsection{System Dependencies} \label{sec:system-dependencies}
As is the case with classical computing systems, dependencies in distributed quantum computing systems limit the ability of the hardware to execute instructions in parallel. While $\rho$ indicates the maximum number of node controllers that can \textit{potentially} be addressed in the same instruction, resource dependencies can prevent different node controllers from \textit{actually} being addressed in the same instruction. To identify these resource dependencies, we can divide the processing of a single instruction into two phases based on the hardware resources that are active.

\begin{enumerate}
    \item \textbf{Issue Phase}: where the central controller is transmitting the instruction over the network interface to the node controller network.
    \item \textbf{Execution Phase}: where the instruction is being executed by the targeted node controller(s).
\end{enumerate}

In the issue phase, the active hardware resource is the network interface. Contention on the network interface constrains parallelism, since the contents of only one instruction can be transmitted by the interface during any given clock cycle. We refer to this type of dependency as an \textit{interface dependency}. On the other hand, in the execution phase, any targeted node controller is an active hardware resource. Consequently, there is a dependency between instructions that target physical qubits controlled by the same node controller. Since the group of physical qubits controlled by a node controller all share the same signal generation hardware in the quantum-classical interface, the node controller can only perform operations on one physical qubit at a time. We refer to this type of dependency as a \textit{node controller dependency}. Based on these two dependency types, we can organize sets of instructions into three different categories.

% As is the case with classical computing systems, dependencies in distributed quantum computing systems limit the ability of the hardware to execute instructions in parallel. While $\rho$ indicates the maximum number of node controllers that can \textit{potentially} be addressed in the same instruction, two types of resource dependencies in distributed quantum systems can prevent different node controllers from \textit{actually} being addressed in the same instruction. At the most fundamental level, there is a dependency between instructions that target physical qubits controlled by the same node controller. Since the group of physical qubits controlled by a node controller all share the same signal generation in the quantum-classical interface, the node controller can only perform operations on one physical qubit at a time. We refer to this type of dependency as a \textit{node controller dependency}. At a higher level, parallelism is also constrained by contention on the network interface, since the contents of only one instruction can be transmitted by the interface during any given clock cycle. We refer to this type of dependency as an \textit{interface dependency}. Based on these two dependency types, we can organize sets of instruction sequences into three different categories.

% Based on these two dependencies, we can divide the execution of a single instruction into two phases based on what hardware resources are used.

\subsubsection{Serial Sequences}
A serial sequence is a sequence of instructions which have node controller dependencies between each other (i.e, the instructions all target the same node controller). As a result, the central controller must wait both for the current instruction's issue phase and execution phase to complete before it can issue the next instruction in the sequence.

\subsubsection{Pipelined Sequences}
A pipelined sequence is a sequence of instructions which have only interface dependencies between each other. Consequently, the central controller only needs to wait for the current instruction's issue phase to complete before it can begin issuing the next instruction to the next target node controller. The condition for a pipelined instruction sequence is that each instruction must target a different node controller. This condition avoids any node controller dependencies between the instructions in a pipelined sequence.

\subsubsection{Parallel Sequences} \label{subsec:parallel_sequence}
A parallel sequence is a sequence of instructions which have neither node controller nor interface dependencies between each other. Due to the absence of dependencies, the central controller can initiate execution of the instructions at the same time. In order to fall into this category, each instruction in the sequence must satisfy the following conditions relative to all other instructions in the sequence.

\begin{itemize}
    \item Each instruction must target a different node controller.
    \item Each instruction must perform the same operation.
    \item Each instruction must have the same instruction parameters (e.g., rotation angle).
\end{itemize}

\noindent
Since the first condition mandates that each instruction in the sequence must target a different node controller, node controller dependencies between the instructions are avoided. Furthermore, since the second and third conditions guarantee that the same opcode and instruction parameters apply to every instruction in the sequence, each node controller can receive its necessary information from a single issued instruction. Therefore, interface dependencies are also circumvented.

\subsection{Distribution Modes} \label{subsec:dist_modes}
\begin{figure}%[h!]
    \centering
    \begin{subfigure}{\columnwidth}
        \centering
        \includegraphics[width=0.64\textwidth]{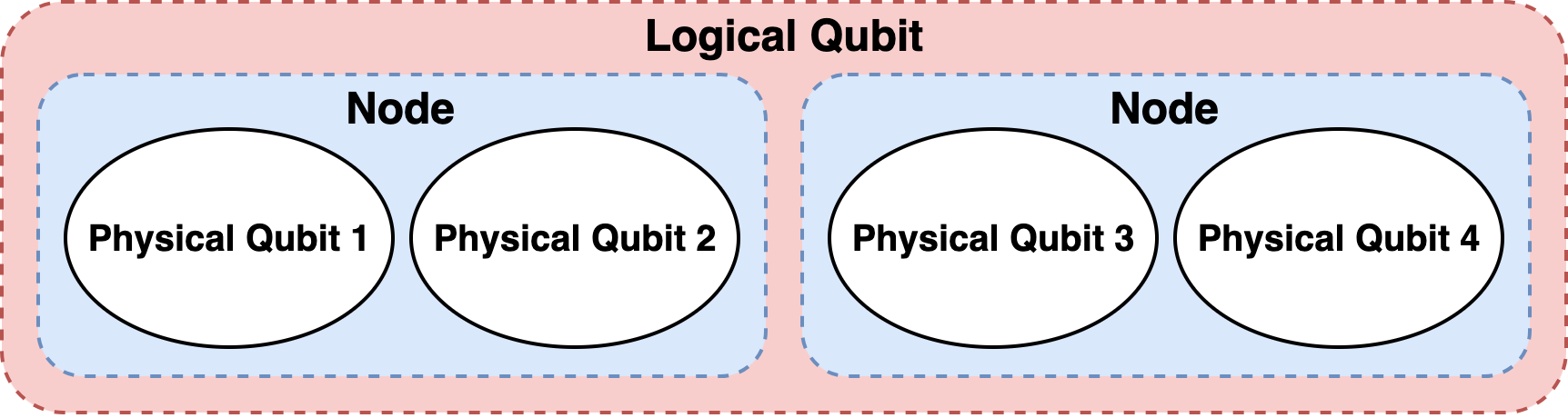}
        \caption{Semi-distributed mode.}
        \label{fig:semi-mapping}
    \end{subfigure}
    \par\bigskip
    \begin{subfigure}{\columnwidth}
        \centering        \includegraphics[width=0.64\textwidth]{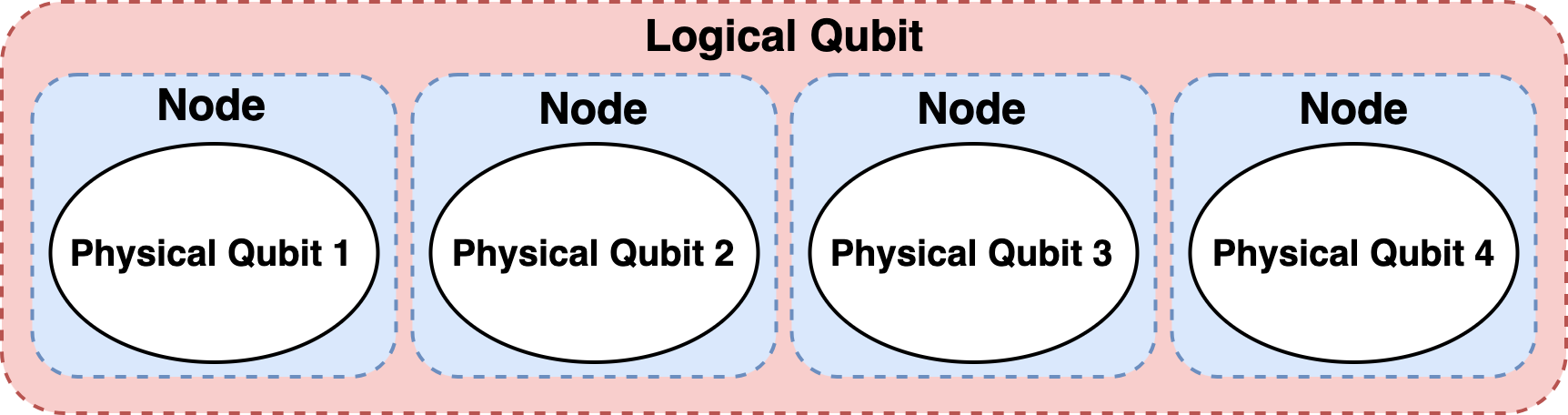}
        \caption{Fully distributed mode.}
        \label{fig:fully-mapping}
    \end{subfigure}
    % \caption{Mappings of a logical qubit onto four physical qubits.}
    \caption{Mapping of a logical qubit to four physical qubit system based on a semi-distributed (top) or fully distributed (bottom) mode. In the semi-distributed mode, two physical qubits are put in the same node. In the fully distributed mode, every physical qubit is put on a different node.}
    % \Description[]{This figure shows the difference in how physical qubits mapped to the same logical qubit are distributed across nodes in the quantum computing system for the semi- and fully distributed modes. The semi-distributed mode distributes a logical qubit across two nodes, each with two physical qubits, while the fully distributed mode distributes the logical qubits across four nodes, each with a single physical qubit.}
    \label{fig:logical-qubit-mapping}
\end{figure}

% To execute logical operations from a quantum circuit we require to know how a logical qubit can be mapped into physical qubits. 
% As mentioned in Section \ref{sec:system-layout}, the quantum circuit representation specifies logical operations on logical qubits. 
Eventually, in order to execute a logical operation from a quantum circuit in hardware, it is necessary to know how logical qubits can be mapped to physical qubits.
% which physical qubit(s) the operation should be performed on. Therefore, we require a method for mapping logical qubits in the quantum circuit to physical qubits in the quantum chip.
Within the context of this paper, we consider two ways of mapping a logical qubit onto four physical qubits.  These mappings are compatible with the distance-2 surface code which is a 
% well-understood 
mechanism for enabling error detection on physical qubits \cite{marques}. It is however important to note that the methodologies that we use are not limited to only a mapping onto four qubits but can be used for any amount of mapped qubits. The first mapping, shown in Fig. \ref{fig:semi-mapping} and referred to as the \textit{semi-distributed mode}, distributes the logical qubit over two separate nodes, where each node contributes two qubits. The second method, shown in Fig. \ref{fig:fully-mapping} and referred to as the \textit{fully distributed mode}, distributes the logical qubit over four separate nodes.

In the semi-distributed mode, since two pairs of physical qubits within a logical qubit are in different nodes, each pair of qubits is controlled by separate node controllers and signal generation hardware. Similarly, in the fully distributed mode, since all four of the physical qubits are in different nodes, they are all separately controlled. Therefore, if the central controller can issue instructions to multiple node controllers simultaneously, there are potential opportunities in both distribution modes for executing operations on physical qubits within the same logical qubit in parallel.

% In both distribution modes, since some physical qubits within a logical qubit are in different nodes, they are controlled by separate node controllers and signal generation hardware. Therefore,

% The system that we are using maps a logical qubit to four physical qubits as shown in Figure \ref{fig:logical-qubit-mapping}. This is a common manner in which logical qubits are mapped, because it allows for a 4 qubit maximally entangled qubit state.

% In the current system, a logical qubit is mapped to four physical qubits as shown in Figure \ref{fig:logical-qubit-mapping}. 
% The manner in which this mapping is done is dependent on the type of distribution mode. We have two distribution modes in our research, either semi or fully distributed modes. We have only two distribution modes, because these make the most logical sense. The fully distributed mode maps all the physical qubits of a logical qubit onto four different nodes, all with their own control hardware. This means that on these qubits, the control hardware execute instructions at the same time, which allows for maximum parallelism. The semi distributed mode maps the physical qubits pairwise to two different nodes, this means that some qubits are controlled by the same hardware. This control hardware will only be capable of execution an instruction on 1 of the qubits at the same time. As a result, some instructions cannot be executed in parallel.

% which results in some instructions to not be executed in parallel.
% Explanation of the mapping from logical to physical qubits

\subsection{Logical Gate Decompositions} \label{sec:logical-gate-operations}
Just as logical qubits must be mapped to a set of physical qubits, logical gates in a quantum circuit must be mapped to a corresponding set of executable physical gate instructions through a process called logical gate decomposition. Here, we consider the decompositions for the logical gate set \{\rx{}, \ry{}, \rz{}, CX\}, since this set comprises a universal set and corresponds to our supporting hardware.  We begin with a set of baseline decompositions, and then we discuss a series of optimizations that can be made to these baseline decompositions to introduce pipelined or parallel execution of physical instructions. It is important to note that the methodology used here is not limited to the above-specified logical gate set and can be used for any gate set. Next to this, the compiler is also capable of transpiling any set of logical gates into the logical gate set used by our system.

% In current quantum algorithm benchmarks, instructions correspond to operations at a logical level. However, to execute these instructions on a quantum system, they must be mapped to their corresponding physical instructions. The decomposition process can vary based on if instructions are parallel, pipelined or sequential. Therefore in this subsection, the different cases of decomposition are discussed. Some gates are a special case that has some additional optimization.
% We have decided to work with a gate set consisting of an \rx{}, \ry{}, \rz{} and CX gate, because this comprises an universal set. 
% Current quantum algorithm benchmarks have instructions that correspond to operations on a logical level. This means that every instruction will have to be mapped to its physical corresponding instruction in order to be executed on a quantum system. The manner in which these decompositions can happen can depend on wether a set of instructions are parallel, pipelined or another specific case.

\subsubsection{Baseline Logical Gate Decompositions} \label{subsubsec:baseline_decomp} %\hfill \\ 
\begin{figure}[]
% \resizebox{\textwidth}{!}{  
\centering
\begin{subfigure}{.21\textwidth}
\centering
  \includegraphics[width=1.1\textwidth]{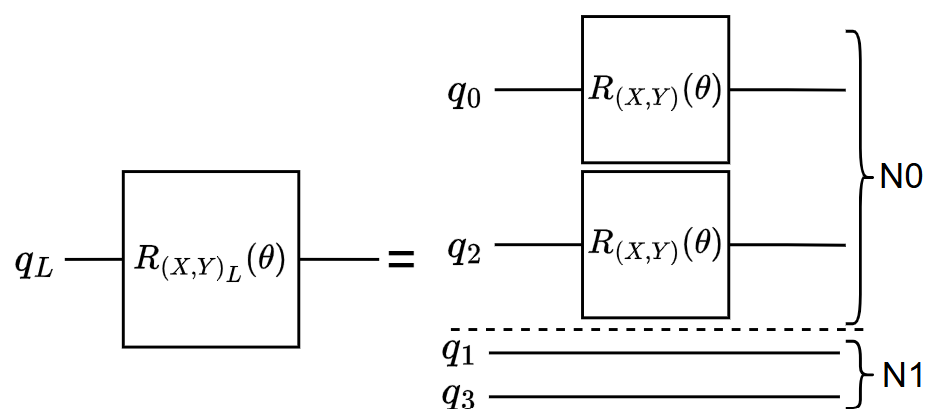}
  \caption{\rx{} or \ry{}.}
  \label{fig:RXY_decomp}
\end{subfigure}
% \par\bigskip
\hspace{0.5cm}
\begin{subfigure}{.21\textwidth}
  \centering
  \includegraphics[width=0.9\textwidth]{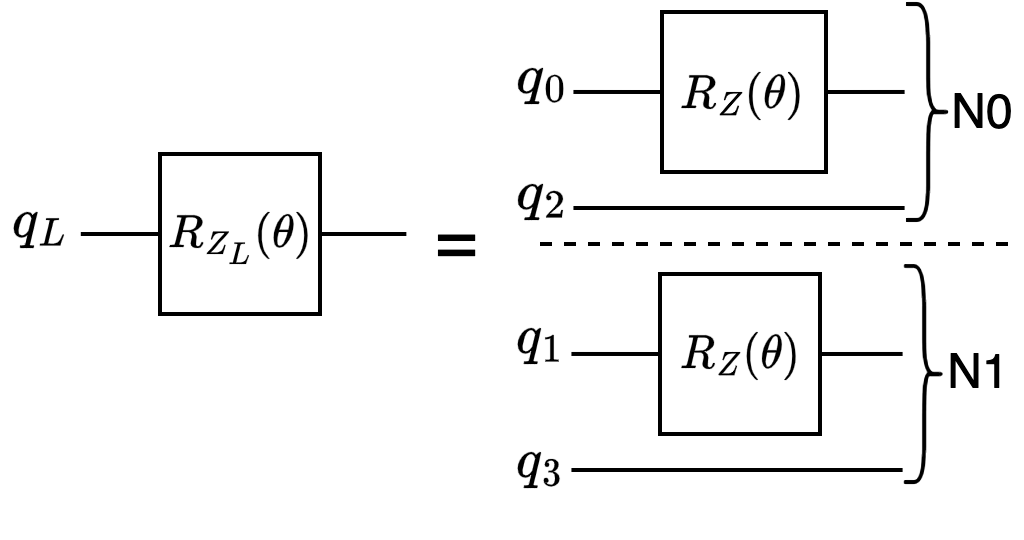}
  \caption{\rz{} gate.}
  \label{fig:RZ_decomp}
\end{subfigure}%\hspace{1cm}
\hspace{1cm}
\begin{subfigure}{.34\textwidth}
\centering
  
  \includegraphics[width=1\textwidth]{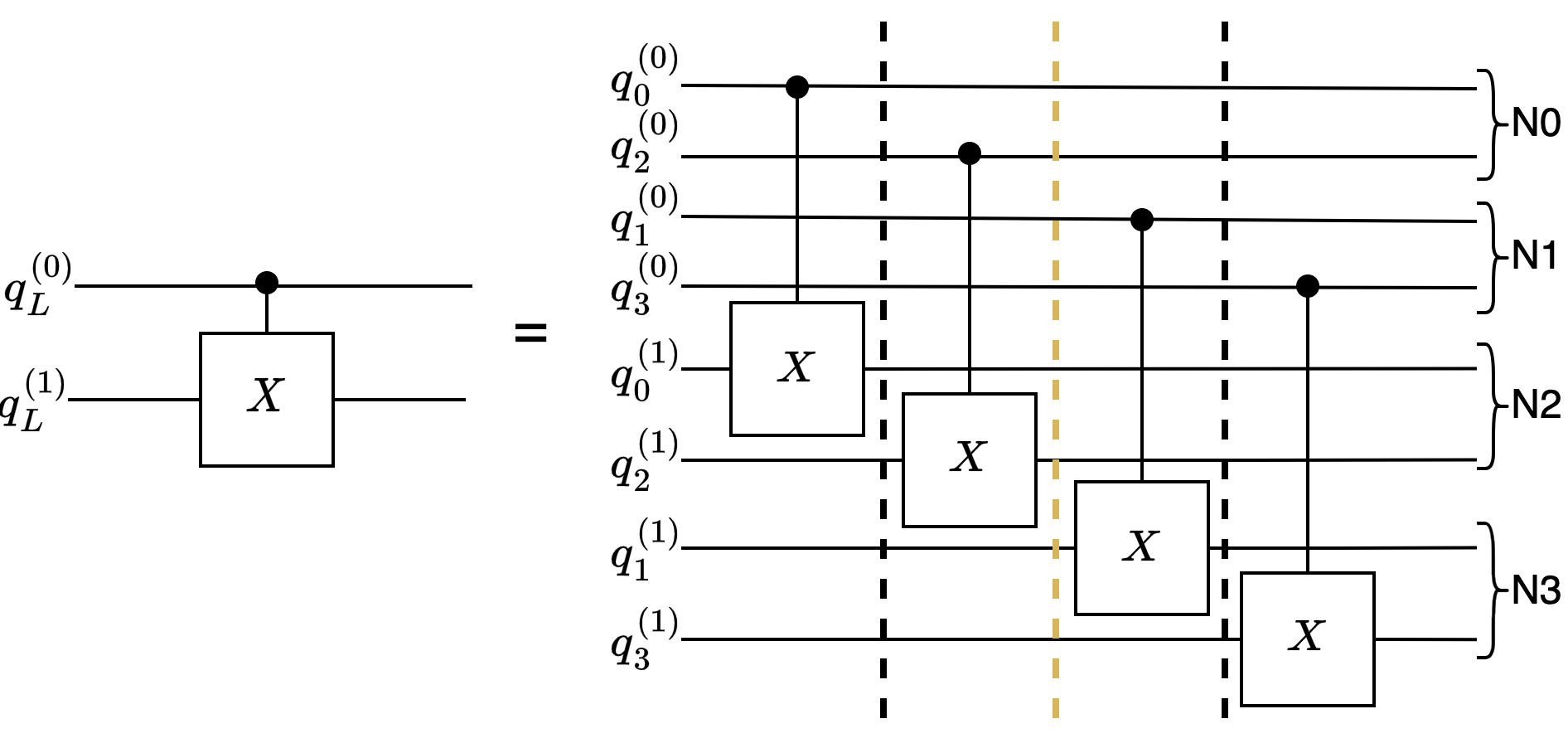}
  \caption{CX gate.}
  \label{fig:cx_basic}
  
\end{subfigure}

% \caption{Baseline decompositions of the logical gates for the semi-distributed mode.}
\caption{A basic decomposition for the Rx, Ry, Rz and CX gates for the semi-distributed mode. The Rx and Ry gates have equivalent decompositions. The CX gate decomposition shows black dotted and orange dotted lines, representing a sequential and pipelined operation respectively.}
% \Description[]{This figure shows three logical to physical decompositions for 4 gates. Figure A presents a decomposition for a rotation X or rotation Y gate gate, where 1 logical operation is decomposed into two physical operations on 1 node. Figure B presents the decomposition for rotation Z gate, where 1 logical operation is decomposed into two physical operations distributed over 2 nodes. Figure C presents a decomposition for the logical CX gate, where 1 logical operation is decomposed into 4 physical cx gates distributed over 4 nodes. }

\label{fig:basic-decomp}
\end{figure}

 The baseline decompositions of the logical gates for the semi-distributed mode are presented in Fig. \ref{fig:basic-decomp}; the logical \ry{} gate decomposition is omitted, since it takes the same form as the logical \rx{} gate decomposition. For each logical gate, the decompositions for the fully distributed mode are the same, but every physical qubit is on a separate node. 
 
\begin{figure*}[h!]
    \centering
    \includegraphics[width = 1\textwidth]{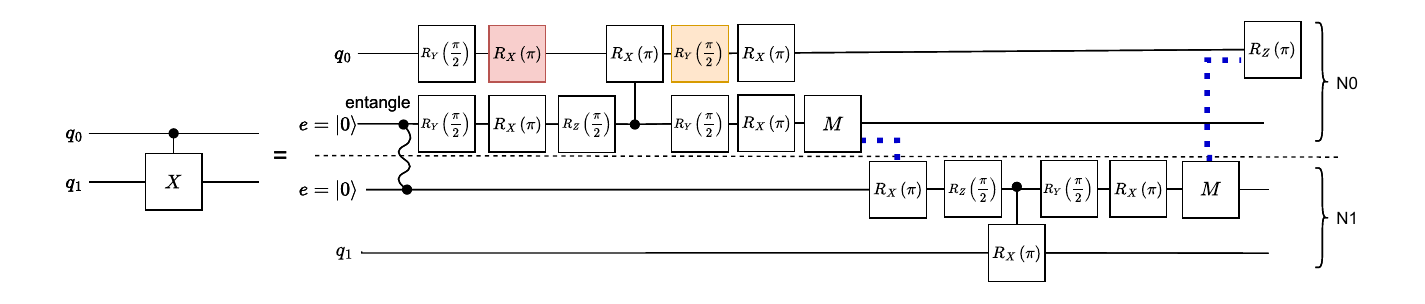}
    % \caption{Full physical CX gate decomposition (adapted from \cite{cx_gate}).}
    \caption{Full physical CX gate decomposition (adapted from \cite{cx_gate}). The first squiggly line represents an entangle operation. The operations presented in red and yellow represent Rx and Ry operations that can potentially be executed in parallel with Rx or Ry operations following the CX gate. The dotted lines coming from the measurement gate connect the measurement value with the controlled operation.}
    % \Description[]{This figure shows the full decomposition of a CX gate between two qubits on different nodes. Performing a cx gate between two gates on distant nodes requires gate teleportation. In this gate teleporation algorithm, a rotation X gate and a rotation Y gate are performed on the control qubit. These rotation gates can be executed in parallel with rotation X and rotation Y gates that follow this cx gate.}
    \label{fig:full_cx_decomp}
\end{figure*}

 The decomposition for a single logical \rx{} or \ry{} gate requires performing two physical instructions using the same node controller (as presented in Fig \ref{fig:basic-decomp} on node N0). Consequently, these instructions comprise a serial sequence due to the node dependency. However, in the fully distributed mode, each physical qubit is in a separate node, so the physical instructions are parallelizable. For the logical \rz{} gate decomposition in Fig. \ref{fig:RZ_decomp}, the physical instructions are performed on different nodes (N0 and N1 in the figure), therefore they can be parallelized in both the semi- and fully distributed modes. Finally, from Fig. \ref{fig:cx_basic}, a logical CX gate is first decomposed into an intermediate representation of four physical CX gates. In the semi-distributed mode, the control and target qubits for the first and second as well as for the third and fourth physical CX gates are found on nodes N0 and N2 respectively N1 and N3, therefore they cannot be parallelized. However, in the fully distributed mode, all physical CX gates can be parallelized as the qubits are all spread over different nodes. It is important to note that the physical CX gates themselves are not directly executed in hardware. Instead, each physical CX gate must be further decomposed into a sequence of instructions as depicted in Fig. \ref{fig:full_cx_decomp}. This final decomposition contains all the instructions needed to perform a physical CX gate between two physical qubits on different nodes.
% which result in no potential for parallelism or pipelining. Their decomposition is presented in Figure \ref{fig:RXY_decomp}. The decomposition of an Rz gate on the other hand is always fully parallel, because it can be performed on different nodes. This decomposition is presented in Figure \ref{fig:RZ_decomp}.

% \textbf{CX gate}

% these CX gates need to be further decomposed onto the physical instruction level, presented in Figure \ref{fig:cx_basic}. However, these intermediate CX gates cannot actually be directly executed on the quantum system. Therefore, they are further decomposed into the sequence shown in Figure \ref{fig:full_cx_decomp}. This final decomposition represents the instructions needed to perform a CX gate between two data qubits on different nodes.

% \begin{figure}
% \centering
  
%   \includegraphics[width=0.5\textwidth]{Figures/cx_seq.PNG}
%   \caption{Basic CX decomposition}
%   \label{fig:cx_basic}
  
% \end{figure}%

% \begin{figure*}[h!]
% % \resizebox{0.5\textwidth}{!}{  
% \centering
% \begin{subfigure}{.5\textwidth}
% \centering
  
%   \includegraphics[width=1\textwidth]{Figures/cx_seq.PNG}
%   \caption{Basic CX decomposition}
%   \label{fig:cx_basic}
  
% \end{subfigure}%
% \begin{subfigure}{.5\textwidth}
%   \centering
%   \includegraphics[width=1\textwidth]{Figures/cx_opt.PNG}
%   \caption{Optimized CX decomposition}
%   \label{fig:cx_opt}

% \end{subfigure}
% % }
% \caption{Decomposition of CX gate}
% \label{fig:decomp-cx}
% \end{figure*}

\subsubsection{Optimized Logical CX Gate Decomposition} \label{subsubsec:optimized_CX_decomp}
\begin{figure}[H]
    \centering
    \includegraphics[width=1\linewidth]{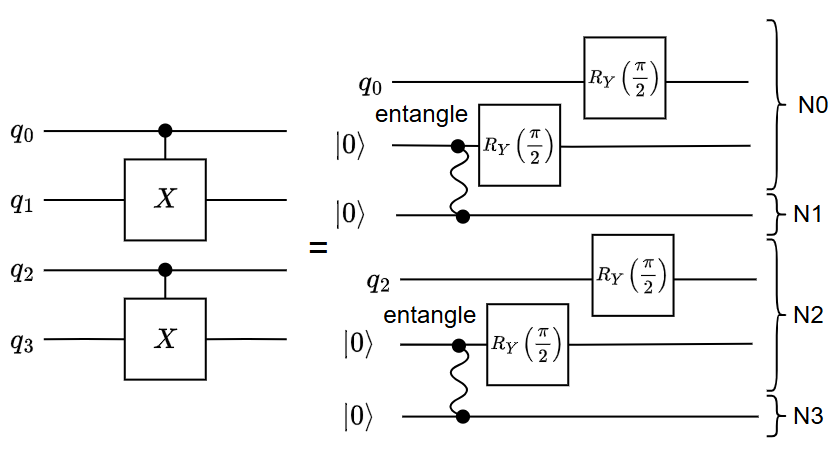}
    % \caption{Pairwise CX gate decomposition sample}
    \caption{A sample of the CX gate presented in Fig. \ref{fig:full_cx_decomp}, showcasing the potential to perform pairwise parallel operations when two logical CX gates can be executed in parallel.}
    % \Description[]{This figure shows the start of the decomposition of two physical CX gates that are in parallel. In the figure, the first three instructions of the decomposition are presented to be executable in parallel.}
    \label{fig:pairwise_decomp}
\end{figure}

\begin{figure}[h!]
  \centering
  \includegraphics[width=0.45\textwidth]{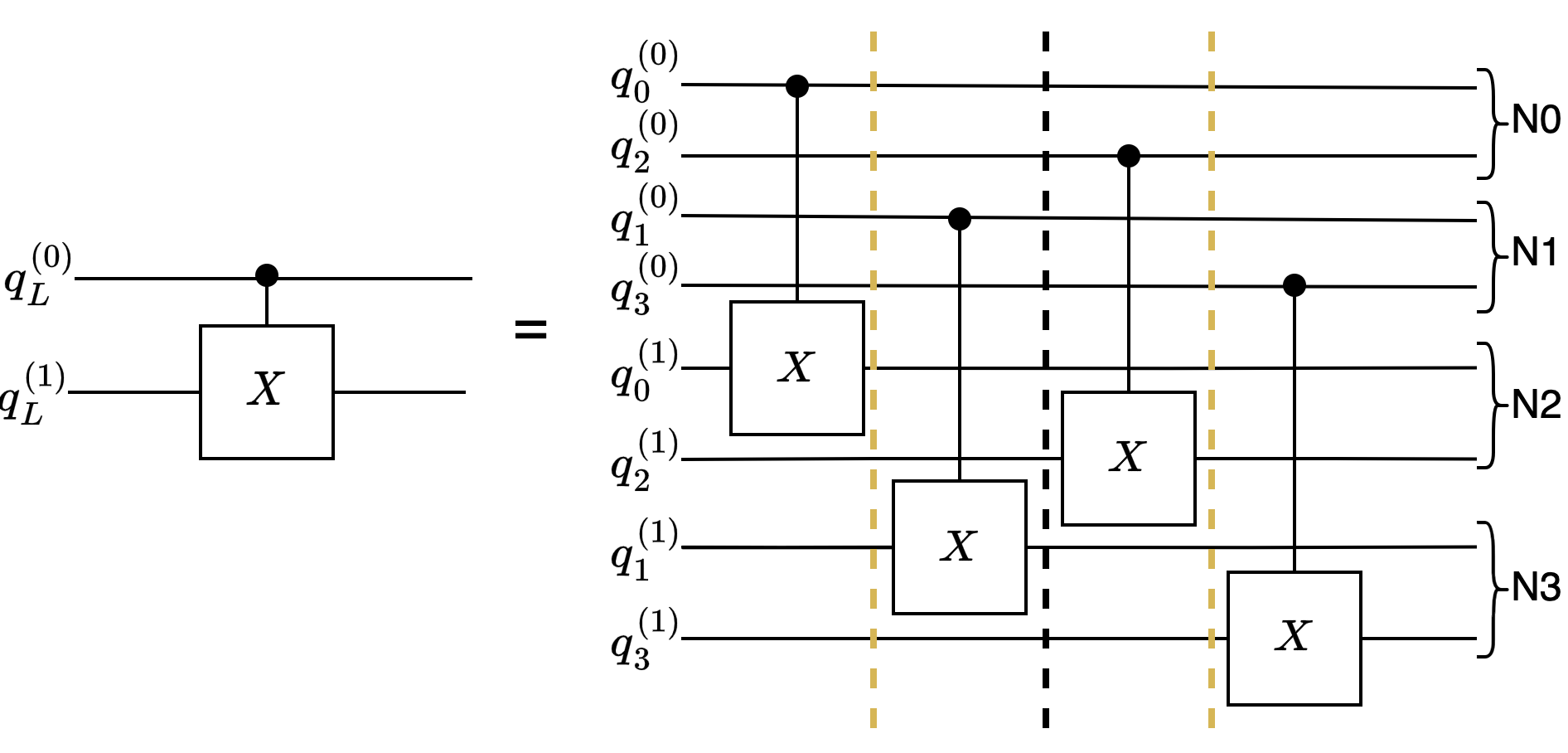}
  % \caption{Optimized CX decomposition.}
  \caption{An optimized decomposition for the CX
gate for the semi-distributed mode. The CX gate decomposition
shows 1 black dotted and 2 orange dotted lines, representing
sequential and pipelined operations respectively. This reduces the amount of sequential instructions by 1 for every CX gate in the parallel sequence compared to Fig. \ref{fig:cx_basic}. }
% \Description[]{This figure shows a decomposition from a logical CX gate into its intermediate representation. The intermediate representation has 4 cx gates that allows for parallel execution. In this case sets of cx gates can be executed in parallel.}
  \label{fig:cx_opt}
\end{figure}

The CX gate decomposition in Fig. \ref{fig:cx_basic} has two sets of serial instructions separated by black dotted lines and one set of parallelizable instructions separated by a yellow dotted line. However, we can decrease the logical gate execution time by rescheduling the physical CX gates according to Fig. \ref{fig:cx_opt}. In this version, there are two sets of parallelizable CX gates (indicated with yellow line) and only one serial set (indicated with black line). When these parallelizable physical CX gates are decomposed into the set of executable instructions from Fig. \ref{fig:full_cx_decomp}, corresponding instructions from each physical CX gate can be scheduled to be pairwise parallel. A sample of the pairwise decomposition is presented in Fig. \ref{fig:pairwise_decomp}.

Furthermore, these decomposed instructions include \rx{} and \ry{} gates, highlighted in red and yellow in Figure \ref{fig:full_cx_decomp}, which can be used to further increase parallel execution.
% These CX gates are then decomposed into a set of instructions which also include an \rx{} and an \ry{} gate on the dataqubits (highlighted with red and yellow in the figure). These instructions can be used to achieve more parallelism.
If a physical CX gate is followed by one or more \rx{} or \ry{} gates that share no node dependency with the CX gate, these gates can be rescheduled in parallel with the \rx{} or \ry{} gate within the CX decomposition.

\subsubsection{Optimized Parallel Sequence Decomposition} \label{subsubsec:optimized_parallel_sequence_decomp}%\hfill\\
\begin{figure*}[h!]
\centering
    \begin{subfigure}{.325\textwidth}
    \centering
      \includegraphics[width=1\textwidth]{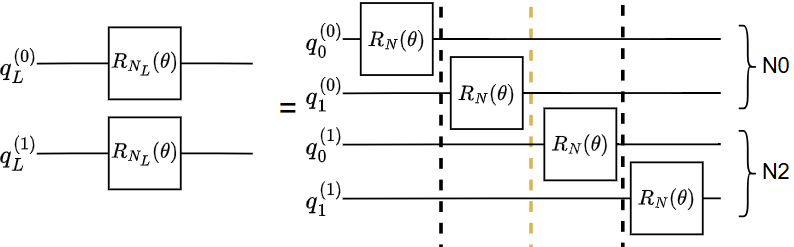}
      \caption{Default decomposition.}
      \label{fig:rx_sequence_not_optimized}
    \end{subfigure}
    \begin{subfigure}{.325\textwidth}
      \centering
      \includegraphics[width=1\textwidth]{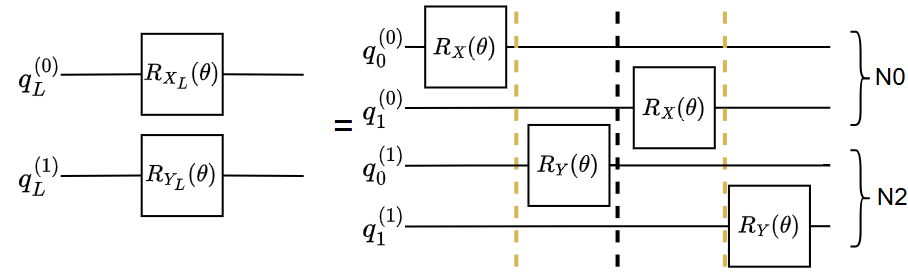}
    \caption{Optimized pipelined decomposition.}
      \label{fig:rx_sequence_optimized}
    \end{subfigure}
\begin{subfigure}{.325\textwidth}
    \centering
    \includegraphics[width = 1\textwidth]{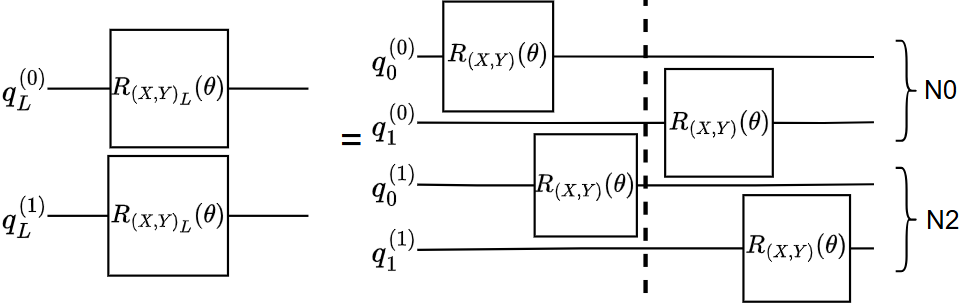}
    % \caption{Pipelined \rx{} and \ry{} gates.}
      \caption{Optimized parallel decomposition.}
    \label{fig:rx_ry_pipe}
\end{subfigure}

% \caption{Optimized rescheduling of decomposed \rx{} or \ry{} instruction sequences.}
\caption{Standard and optimized decomposition for \rx{} and \ry{} gates in a parallel/pipelined sequence.  The decomposition shows black dotted and orange dotted lines, representing
sequential and pipelined operations respectively. The default decomposition (a) has 2 sequential and a pipelined instruction per logical instruction. The pipelined decomposition (b) reduces the amount of sequential instructions by 1 per logical instruction. The parallel decomposition (c) additionally removes the pipelined instructions.}
% \Description[]{This figure shows three different decompositions of two parallel rotation x or y gates. Figure A shows the default decomopsition, in this case two rx gates are decomposed onto 4 physical rotation gates on two nodes, however the first two and last two operations are performed in parallel, while the middle two operations can be performed in pipeline. In figure B and C, the operations are re-ordered to allow for more parallel or pipelined execution. In this case the first and last two operations are performed in parallel or pipeline, while the sets of operations need to be sequential. The instructions are executed in parallel if they are the same instruction and they are performed in pipeline otherwise.}
\label{fig:optimize_ryx_seq}
\end{figure*}

Optimizations are also possible when considering the decomposition of parallel sequences of logical gates. For instance, the baseline decompositions for parallel sequences of logical \rx{} gates is shown in Fig. \ref{fig:rx_sequence_not_optimized}; the same decomposition applies for logical \ry{} gates. This decomposition yields two pairs of serial physical gates. However, as shown in Fig. \ref{fig:rx_sequence_optimized}, the instruction sequence run time can be reduced by alternatively decomposing the logical sequence into two parallel sequences of physical instructions. 

% Furthermore, similar to the decomposition of a single logical CX gate, a parallel sequence of logical CX gates can be decomposed such that corresponding intermediate CX gates in each composition are executed in parallel. The corresponding physical gates of the intermediate CX gates are then executed in parallel with the parallel sequence length equal to the parallel intermediate CX gate length.

Furthermore, like the decomposition of a single logical CX gate, a series of parallel logical CX gates can be decomposed so that their intermediate CX gates are executed in parallel. As a result, the physical gates, corresponding to the intermediate CX gate decomposition, are also executed simultaneously. 

At last, decomposition optimizations for parallel sequences of \rz{} gates are unnecessary since the decomposition of logical \rz{} gates already results in parallel instructions.
% individual logical \rz{} gates are already decomposed into two parallel physical instructions.

% Sometimes, a sequence of parallel instructions can be decomposed more intelligently than the standard approach, resulting in increased parallelism. This is for instance the case for \rx{} and \ry{}, but not for \rz{} because the \rz{} gates are already decomposed into two parallel physical instructions. The \rx{} and \ry{} gates can be improved because they are initially decomposed into two physical \rx{} and \ry{} gates on the same node. Unfortunately, these two physical operations are serial and cannot be optimized. However, a sequence of \rx{} and \ry{} gates can be further decomposed into two parallel sequences of physical instructions, maintaining the same length as the logical parallel sequence. Refer to Figure \ref{fig:optimize_ryx_seq} for an illustration of this optimization.

% The \rz{} gate is already decomposed into 2 parallel physical instructions, so no optimization is needed. Any parallel sequence of \rz{} gates remains fully parallel.

% The CX gate is decomposed into a sequence of 17 instructions. Due to the nature of the decomposition of the CX gates, one can reschedule the decomposition of a single logical CX gate into pairwise parallel physical instructions. The decomposition algorithm for a CX gate is presented in Algorithm \ref{alg:decomp_cx_parallel}.

% \begin{algorithm}[H]
% \caption{Decomposition of CX gate in parallel}
% \label{alg:decomp_cx_parallel}
% \begin{algorithmic}
% \Require List of parallel qubits: \textbf{Parallel\_qubits} 
% \For {Every instruction in CX decomposition}
% \For {Every qubit in \textbf{Parallel\_qubits}}
% \State Perform instruction on qubit
% \EndFor
% \EndFor
% \end{algorithmic}
% \end{algorithm}

% \begin{algorithm}[h!]
% \caption{Parallelized Decomposition of Logical CX gate}
% \label{alg:decomp_cx_parallel}
% \begin{algorithmic}
% \Require List of parallel qubits, $Q$, and CX gate sequence, $G$.
% \For {each $g \in G$}
% \For {all $q \in Q$}
% \State Perform $g$ on $q$
% \EndFor
% \EndFor
% \end{algorithmic}
% \end{algorithm}

\subsubsection{Pipelined Parallel Sequence Decomposition} \label{subsubsec:optimized_pipelined_sequence_decomp}%\hfill \\
Once all the parallel sequences have been identified, there is one final optimization to consider. When two parallel sequences follow each other, and they do not share any node dependencies, there is potential for pipelining the execution of the two parallel sequences to further reduce execution time. 

Consider a pipelined sequence consisting of one logical \rx{} gate and one logical \ry{} gate. By default, the decomposition of this sequence would result in two serial physical \rx{} gates followed by two serial physical \ry{} gates. However, by combining these pipelined gates with decompositions similar to those in Fig. \ref{fig:optimize_ryx_seq}, we obtain the decomposition shown in Fig. \ref{fig:rx_ry_pipe}. The decomposition reduces run time by pipelining pairs of physical \rx{} and \ry{} gates.

% an improvement can be obtained by decomposing the logical gates according to Figure \ref{fig:rx_ry_pipe}. In this diagram, the yellow line represents a pipelined instruction, while the black dotted line corresponds to a serial instruction. By combining these pipelined instructions with the parallel decompositions shown in Figure \ref{fig:optimize_ryx_seq}, we obtain a parallel sequence of \rx{} gates pipelined with a parallel sequence of \ry{} gates.

% \begin{figure}[h!]
%     \centering
%     \includegraphics[width = 0.45\textwidth]{Figures/rx_ry_pipe.PNG}
%     \caption{Pipelined \rx{} and \ry{} gates.}
%     \label{fig:rx_ry_pipe}
% \end{figure}

\section{Hardware Architecture Design} \label{Sec:Hardware_design}
This section discusses our hardware architecture design for distributed quantum systems that strikes a balance between increased instruction throughput and increased single-instruction latency. In Sec. \ref{sec:addressing-methods}, we observed competing design constraints for addressing node controllers in instructions. In the SISD model, where instructions contain unique, compact ID-encoded addresses, single-instruction latency is minimized at the cost of low throughput. On the other hand, in the SIMD model, throughput is increased by using larger bitmap-encoded addresses, but this comes at the cost of higher single-instruction latency. Since both low throughput and high single-instruction latency impact the maximum achievable performance of the quantum computing system, the best hardware design will be one that can balance increasing throughput without significantly impacting single-instruction latency.

\begin{figure*}
    \centering
    \includegraphics[width=0.7\textwidth]{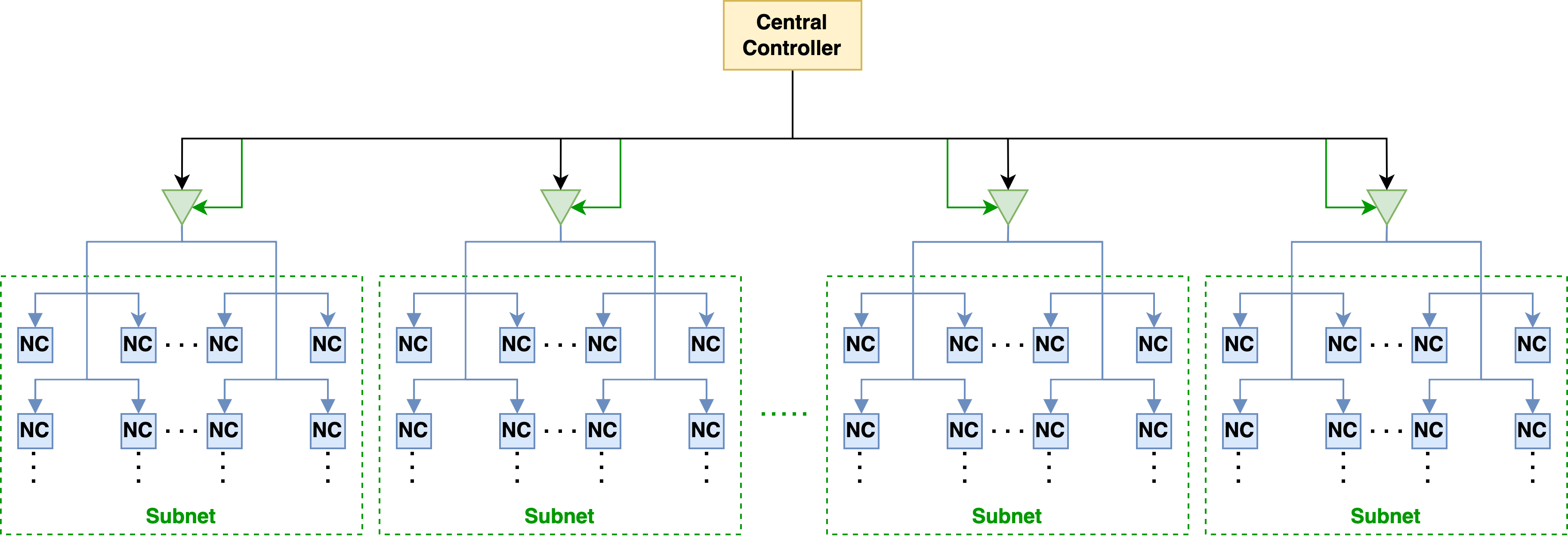}
    \caption{Proposed two-level hierarchical network design. At the first level, a subnet address is used to select which subnet(s) to target with an instruction (represented by the green paths). At the second level, a node controller address is used to select which node controller(s) to target (represented by the blue paths).}
    % \Description[]{The two-level hierarchical design partitions the network of node controllers into groups, called subnets, and adds a new field to each instruction called the subnet address. With this design, a node controller within any subnet can be uniquely identified through a combination of its subnet and node controller addresses.}
    \label{fig:two-level}
\end{figure*}

To obtain the balance between increased throughput and decreased single-instruction latency, we require a scheme for encoding addresses in quantum instructions that is capable of using a compact address representation to target multiple node controllers in parallel. Both design goals can be achieved by dividing address decoding into cascaded stages. Our proposed hardware architecture design implements this cascaded address decoding using a two-level hierarchical network, depicted in Fig. \ref{fig:two-level}, in which node controllers (marked as NC) are partitioned into smaller groups called \textit{subnets} (marked as green boxes). For a network size of $N$ node controllers, partitioning the network into $M$ subnets yields subnets with $K = \frac{N}{M}$ node controllers. Node controllers can then be targeted by an instruction using a combination of two smaller addresses: a first-level address which selects one or more subnets in the network (decoder hardware marked as green triangles), followed by a second-level address which selects one or more node controllers within the subnet(s) being targeted. These two addresses are referred to as the \textit{subnet address} and \textit{node controller address}, respectively. 

We now present an analysis of this hardware architecture design proposal. We begin by discussing high-level design categories for encoding the subnet and node controller addresses, which we refer to as \textit{address encoding schemes}. Within each of these schemes, we also determine how to describe and compare between specific hardware implementations, which we refer to as \textit{addressing modes}. Subsequently, we assess the best strategy per address encoding scheme for mapping logical qubits to subnets.

\subsection{Hardware Design Options} \label{sec:design-options}
For each of the subnet and node controller addresses, we can consider whether to use an ID encoding or a bitmap encoding, yielding four possible address encoding schemes. However, we disregard the case of using an ID encoding for both the subnet and node controller addresses, since this reverts to the SISD model, where the upper bits of the address correspond to the subnet address and the lower bits correspond to the node controller address. Consequently, there are three address encoding schemes we consider.

\begin{enumerate}
    \item \textbf{Subnet ID, Node Controller Bitmap:} enables \textit{intra-subnet parallelism}, where any combination of node controllers within the same subnet can be addressed by the same instruction. For convenience, we denote this encoding scheme as \textit{subID\_ncBIT}.

    \item \textbf{Subnet Bitmap, Node Controller ID:} enables \textit{inter-subnet parallelism}, where the same node controller within any combination of subnets can be addressed by the same instruction. We denote this encoding scheme as \textit{subBIT\_ncID}.

    \item \textbf{Subnet Bitmap, Node Controller Bitmap:} enables parallelism both within and between subnets. However, a single instruction is only able to address the same group of node controllers within any combination of subnets. We denote this encoding scheme as \textit{subBIT\_ncBIT}.
\end{enumerate}

\noindent
To simplify our discussion of specific hardware implementations within each of these address encoding schemes, we first formulate a way of representing the subnet and node controller address widths using the network dimensions of the hierarchical network design. The network dimensions can be described in terms of two parameters: the number of subnets into which the network is partitioned, $M$, and the number of node controllers on each subnet, $K$. For a network of $N$ node controllers, the relationship $N = M \cdot K$ always holds. The widths of the subnet and node controller addresses can be determined based on these network dimensions. Considering the subID\_ncBIT address encoding scheme, the width of the ID-encoded subnet address, $W_S$, and the width of the bitmap-encoded node controller address, $W_{NC}$, can be calculated as $W_S = \log_2(M)$ bits and $W_{NC} = N$ bits, respectively. The address widths for the subBIT\_ncID address encoding scheme are symmetrical to subID\_ncBIT, since now $W_S = M$ bits and $W_{NC} = \log_2(N)$ bits. Finally, when choosing for the subBIT\_ncBIT address encoding scheme, the address widths are $W_S = M$ bits and $W_{NC} = N$ bits. Using this notation, we can uniquely describe the addressing mode corresponding to a particular hardware design using its $\left( W_S, W_{NC} \right)$ address width pairs.

% For each address encoding scheme, multiple addressing modes are possible based on how the node controller network is partitioned into subnets. The network dimensions can be described in terms of two parameters: the number of subnets into which the network is partitioned, $M$, and the number of node controllers on each subnet, $K$. For a network of $N$ node controllers, the relationship $N = M \cdot K$ always holds. The widths of the subnet and node controller addresses can be determined based on these network dimensions. 

According to the hardware design goals, the key figures of merit for evaluating an address encoding scheme are the compactness of the address representation and the number of node controllers that can be addressed in parallel. In all three address encoding schemes, the cumulative width of the subnet and node controller addresses, $W_S + W_{NC}$, for a given network size is always less than the width of the address used in the SIMD design for a single-level network: $W = M \cdot K$ bits. Furthermore, the reduction in address size provided by each encoding scheme relative to the single-level, SIMD design increases as the size of the network grows. At the same time, the maximum achievable throughput in each address encoding scheme is higher than the single-level, SISD network's throughput of $1$ instruction per cycle. In the subID\_ncBIT case, the maximum achievable throughput is $K$ instructions per cycle, while for the subBIT\_ncID and subBIT\_ncBIT encoding schemes, it is $M$ and $MK$ instructions per cycle, respectively. 

% Both of these aspects are key figures of merit for validating the scalability of a particular hardware design.

Within an address encoding scheme, it is useful to relate addressing modes back to the performance tradeoff parameters $\rho$ and $\delta$ defined in Sec. \ref{sec:addressing-methods} in order to quantitatively evaluate different addressing modes relative to each other. $\rho$ corresponds to the maximum number of node controllers that can be simultaneously addressed by a single instruction. Therefore, in the subID\_ncBIT address encoding schemes, $\rho = W_{NC} = K$, the number of node controllers per subnet. In the subBIT\_ncID address encoding scheme, $\rho = W_{S} = M$, so it corresponds to the number of subnets. Lastly, in the subBIT\_ncBIT address encoding scheme, all node controllers in the network can be addressed simultaneously if all bits in both the subnet and node controller bitmap-encoded addresses are set. Consequently, $\rho = N$, the total number of node controllers in the network.

 % We can also relate each addressing mode back to the parameters $\rho$ and $\delta$ defined in Section \ref{sec:addressing-methods}. 
 
 The value of $\delta$ for each addressing mode similarly depends on the mode's address widths, but it is also dependent on the number of data wires, $L$, used in the network interface between the central controller and node controllers. When the cumulative address width of an addressing mode exceeds the number of data wires in the network interface, then the address bits will need to be transmitted over multiple clock cycles. Assuming that the address in the SISD execution model is transmitted in a single cycle, Eq. \ref{eq:delta-calc} captures the additional overhead in address transmission for a particular addressing mode. The first term of the piecewise function covers the special case where the cumulative address width is an exact multiple of the number of data wires in the network interface.

 \begin{equation}
     \delta = \begin{cases}
        \left \lfloor \frac{W_s + W_{NC}}{L} \right \rfloor - 1, & \text{if } W_s + W_{NC} = 0 \bmod L \\ \\
        \left \lfloor \frac{W_s + W_{NC}}{L} \right \rfloor, & \text{if } W_s + W_{NC} \neq 0 \bmod L \label{eq:delta-calc}
    \end{cases}
 \end{equation}

\subsection{Subnet Organization} \label{sec:subnet-organization}
 As mentioned in the previous section, the hardware's ability to facilitate parallelism across node controllers is restricted in different ways based on the address encoding scheme used. In the subID\_ncBIT address encoding scheme, hardware support for parallelism is restricted to node controllers within the same subnet, while for the subBIT\_ncID address encoding scheme, hardware support for parallelism is limited to node controllers across different subnets. Finally, hardware designs within the subBIT\_ncBIT address encoding scheme can only support parallelism between the same groups of node controllers across different subnets.

 Since the primary source of instruction-level parallelism in distributed quantum computing systems comes from repeated instructions performed within the same logical qubit, the subnets within each address encoding scheme should be organized such that the hardware is always capable of supporting parallel instructions within a logical qubit. To accomplish this for the subID\_ncBIT address encoding scheme, we organize subnets such that the set of node controllers used to control the same logical qubit are always assigned to the same subnet. By contrast, for the subBIT\_ncID encoding scheme, the node controllers are always assigned to different subnets. By organizing the subnets in an encoding-scheme-specific manner, it is expected that both encoding schemes should perform equally well, since the parallelism within subnets in the subID\_ncBIT scheme is entirely replaced by parallelism across subnets in the subBIT\_ncID scheme. 
 
\begin{figure*}[h!]
    \centering
    \begin{subfigure}[b][][b]{0.323\textwidth}
        \includegraphics[width=\textwidth]{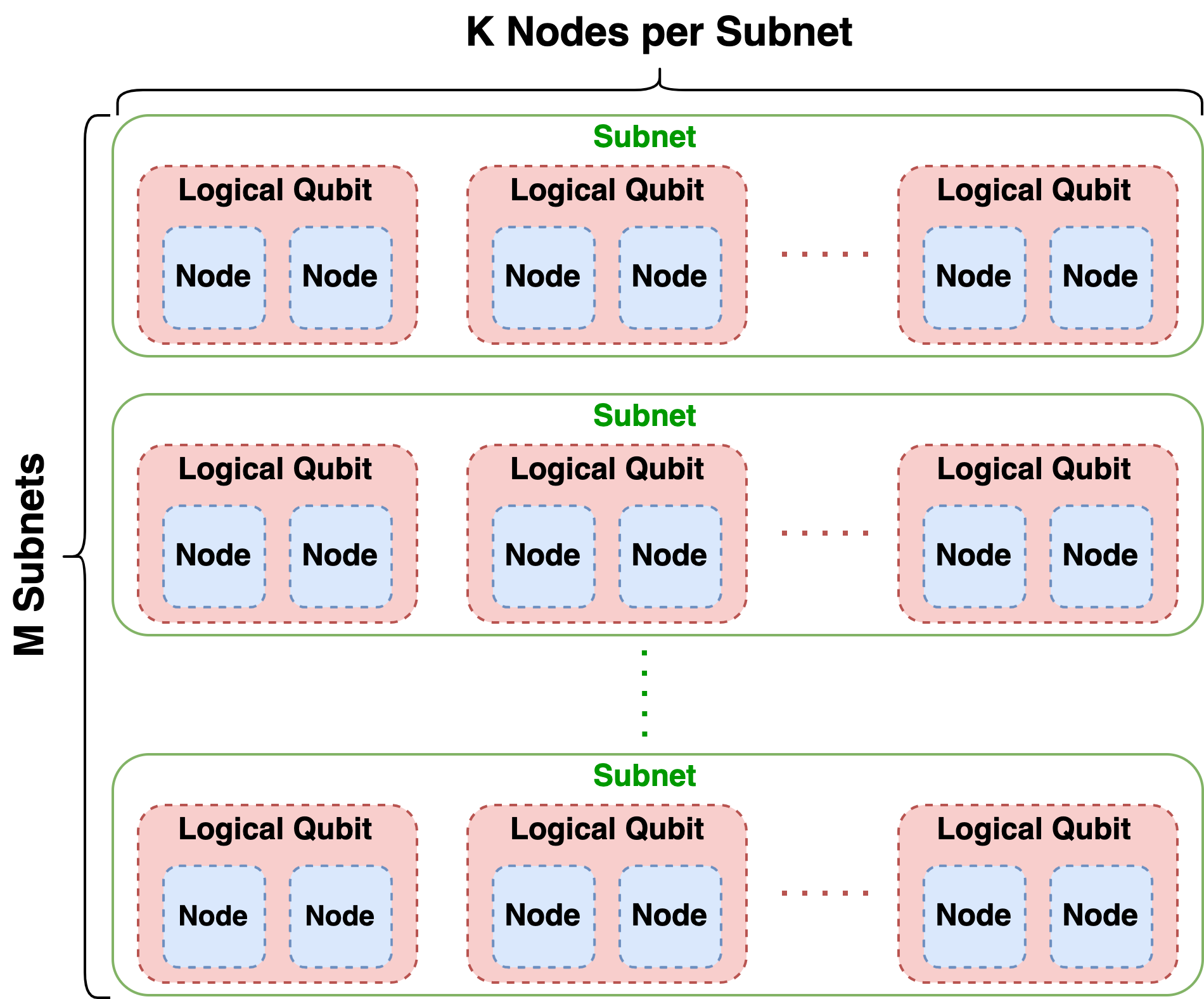}
        \caption{subID\_ncBIT and subBIT\_ncBIT}
    \end{subfigure} \hspace{1cm}
    \begin{subfigure}[b][][b]{0.379\textwidth}
        \includegraphics[width=\textwidth]{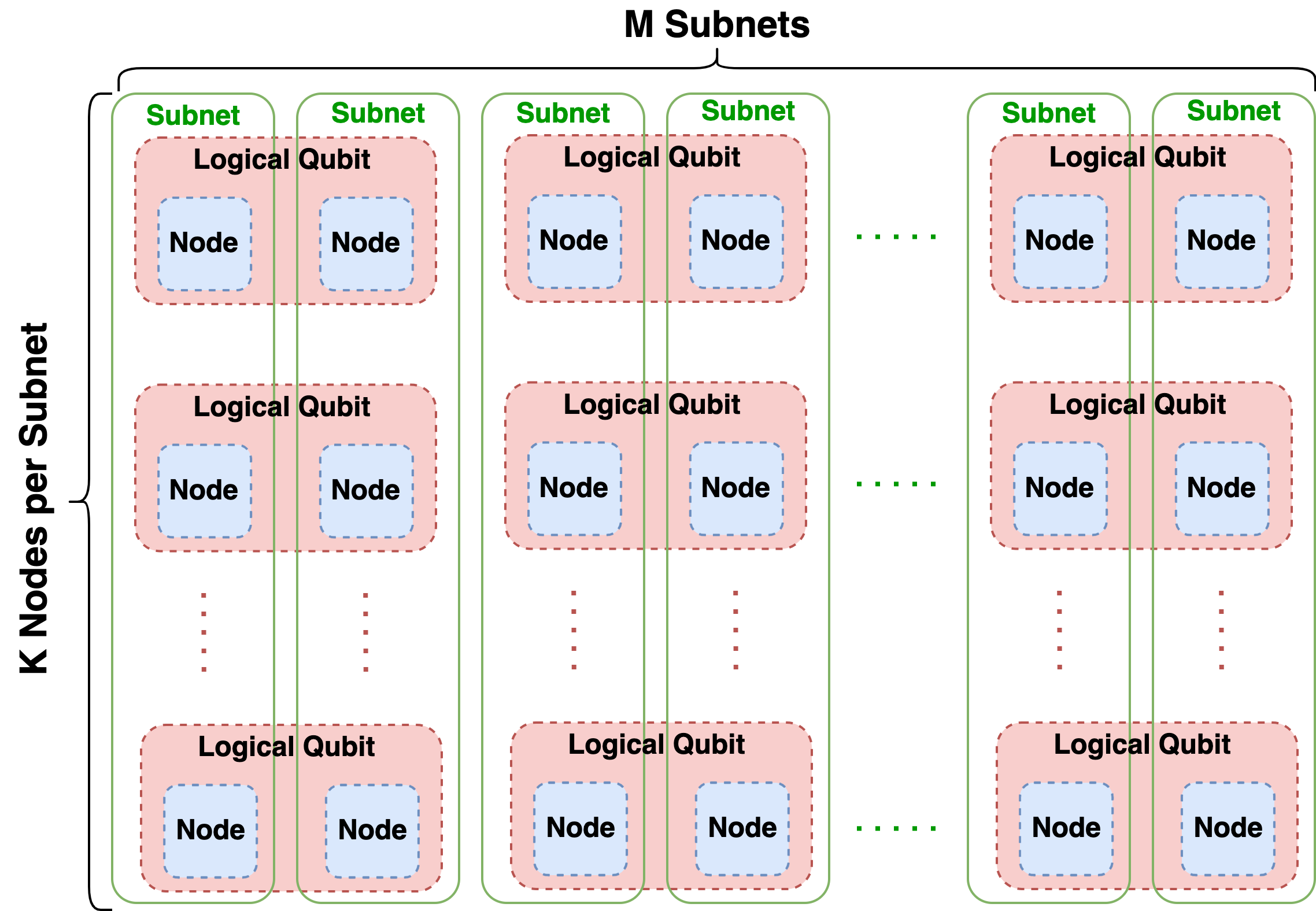}
        \caption{subBIT\_ncID}
    \end{subfigure}
    \caption{Subnet organization per encoding scheme over the node controller network for the semi-distributed mode. (a) In both the subID\_ncBIT and subBIT\_ncBIT encoding schemes, nodes containing physical qubits from the same logical qubit are assigned to the same subnet. (b) In the subBIT\_ncID, nodes containing physical qubits from the same logical qubit are assigned to different subnets.}
    % \Description[]{This figure shows the difference between how the subID\_ncBIT and subBIT\_ncBIT encoding schemes organize nodes into subnets compared to the subBIT\_ncID encoding scheme.}
    \label{fig:subnet_distribution}
\end{figure*}

 For the subBIT\_ncBIT address encoding scheme, while parallelism is possible both within and across subnets using the third address encoding scheme, parallelism across subnets is more constrained. Therefore, as with the subID\_ncBIT encoding scheme, we assign node controllers used to control the same logical qubit to the same subnet. The subnet organization for each of the three address encoding schemes is depicted in Fig. \ref{fig:subnet_distribution}.

Based on these encoding-specific subnet organizations, we can derive functions for determining which subnet a physical qubit is in. Eventually, these will be useful for the compiler when it determines how best to reschedule program instructions for addressing modes within the subID\_ncBIT and subBIT\_ncID encoding schemes. For these two encoding schemes, the subnet corresponding to a specific physical qubit can be calculated using Eq. \ref{eq:subnet_subnetID} and Eq. \ref{eq:subnet_subnetbit}, respectively. In these equations, $q_i$ denotes the physical qubit's index, $N_q$ denotes the number of physical qubits per node, and $\rho$ denotes the parallelizability factor. The term $q_i/N_q$ represents the ID of the node controller to which the physical qubit is assigned.
 \begin{equation}
    S(q_i) = \left \lfloor \frac{\left \lfloor \frac{q_i}{N_q} \right \rfloor}{\rho} \right \rfloor
    \label{eq:subnet_subnetID}
\end{equation}
\begin{equation}
    S(q_i) = \left \lfloor \frac{q_i}{N_q} \right \rfloor \bmod \rho
    \label{eq:subnet_subnetbit}
\end{equation}
% \begin{figure*}[h!]
%     \centering
%     \begin{subfigure}[b][][b]{0.43\textwidth}
%         \includegraphics[width=\textwidth]{Figures/subnet-dist-same.png}
%         \caption{subID\_ncBIT/subBIT\_ncBIT encoding schemes.}
%     \end{subfigure}
%     \begin{subfigure}[b][][b]{0.505\textwidth}
%         \includegraphics[width=\textwidth]{Figures/subnet-dist-diff.png}
%         \caption{subBIT\_lcID encoding scheme.}
%     \end{subfigure}
%     \caption{Subnet organization over the node controller network for the semi-distributed mode.}
%     \label{fig:subnet_distribution}
% \end{figure*}

\section{Compiler Design} \label{sec:Compiler_design}
The hardware architecture design presented in Sec. \ref{Sec:Hardware_design} allows us to perform operations in parallel on our distributed quantum system. However, quantum algorithms are not inherently organized for parallel execution. Therefore, we need a system capable of reordering quantum circuit instructions to increase parallel execution potential without altering the circuit's functionality. 
% It is however not usual that a quantum algorithm is organized in a way that can be executed in parallel. Therefore a system needs to be added that is capable of ordering a quantum algorithm optimally, without changing the functionality of the program. 

\begin{figure}[h!]
    \centering
    \includegraphics[width=0.3\linewidth]{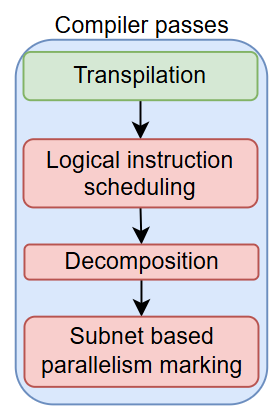}
    \caption{A stacked overview of the compiler passes. The pass encoded in green is already part of Qiskit, while the passes encoded in red are direct contributions of this paper. The arrows represent the order in which the passes are called.}
    % \caption{Compiler passes}
    % \Description[]{This figure shows the 4 passes that comprise the compiler. The compiler uses a transpilation tool, created by qiskit, to decompose instructions into the native gateset. After that the instructions are put through a pass that performs logical instruction scheduling. Then the logical insturctions are decomposed into physical instructions in a optimized manner. At last, the instructions are marked for parallelism based on the type of encoding scheme.}
    
    \label{fig:compiler_passes}
\end{figure}

In this section, we introduce a compiler, based on the Qiskit compiler, with four main functionalities. While the first functionality is already part of Qiskit, we have added the other three. The passes are described below and the general structure is presented in Figure \ref{fig:compiler_passes}.

The compiler can:
\begin{enumerate}
    \item \textbf{Transpile Instructions:} Convert a set of logical instructions into the gate set used by a specific quantum system.
    \item \textbf{Schedule Logical Instructions:} Change the order of logical instructions to increase the amount of logical instructions which can be executed in parallel.
    \item \textbf{Decompose Logical Instructions:} Transform logical instructions into physical instructions executable on specific quantum hardware.
    \item \textbf{Define Subnet-Based Parallelism:} Mark instructions which can be executed in parallel based on the underlying hardware.
    % Perform hardware-dependent parallel sequence marking. 
    The compiler is aware of the encoding scheme and the number of subnets used in the hardware design.
    % Consider subnets (as presented in Section \ref{subsec:encoding_schemes}) to determine the achievable parallelism on a given system.
\end{enumerate}
The first functionality (colored green) leverages Qiskit’s transpiler, which uses the basis gates of the target quantum system. For our case, we consider a native gate set of rotation gates in the x, y, and z bases, as well as the CX (controlled-X) gate.
The three added functionalities (colored red) of the compiler are implemented in separate passes described in Sec. \ref{subsec:Scheduling_pass}, Sec. \ref{subsec:Decomposition_pass}, and Sec. \ref{subsec:Subnet_pass} respectively. 

\subsection{Scheduling Pass}\label{subsec:Scheduling_pass}
The scheduling pass aims to maximize the number of logical instructions that can be executed in parallel. Using a dependency graph, as detailed in Alg. \ref{alg:scheduling}, the pass reorders the instructions.  The algorithm iterates through every instruction and checks if there is any dependency with the previous instructions. Every instruction will then be grouped based on their dependencies.

Alg. \ref{alg:dependency_determination} is employed to determine whether two instructions have a node controller dependency. During this process, the pass identifies instructions within the quantum algorithm that are independent, meaning their execution does not depend on each other. These independent instructions are then grouped together, while ensuring the program's functionality remains unchanged.
% The pass identifies non-dependent instructions within the quantum algorithm and groups them together. Because these instructions are independent, they can be grouped without altering the program's functionality. 
These instructions are then sorted based on their opcodes and parameters, as parallel execution requires identical opcodes and parameters (as discussed in Sec. \ref{sec:design-options}). Additionally, the scheduler automatically arranges instructions for pipelined execution.

\begin{algorithm}[h!]
\caption{Dependency Graph-Based Scheduling}\label{alg:scheduling}
\begin{algorithmic}
\REQUIRE The list of logical instructions, $L$
% \STATE Determine the list of logical instructions: \textbf{instr\_logical}
% \STATE $D \leftarrow$ Create empty set of dependency lists
\STATE /* Initialize dependency list with first logical instruction */
\STATE $d_0 \leftarrow \{l_0\}$
\STATE /* Initialize dependency graph with first dependency list */
\STATE $D \leftarrow \{d_0\}$
\STATE /* Initialize dependency parameter */
\STATE $DP \leftarrow 0$
\FOR {each $l_i \in L$}
\FOR {each $d_j \in D$}
\IF {\textit{hasDependency}($l_i$, $d_j$)}
\IF {$d_{j+1} = \emptyset$}
\STATE /* Create next dependency list */
\STATE $d_{j+1} \leftarrow \{ \}$
\ENDIF
% \STATE \textit{append}($l_{i}, d_{j+1}$)
\STATE $l_{i}.append(d_{j+1})$
\ELSE
\STATE $DP \leftarrow 1$
\ENDIF
\ENDFOR
\IF {$DP == 0$}
% \STATE \textit{append}($l_i, d_0$)
\STATE $l_{i}.append(d_{0})$

\ENDIF
\ENDFOR
\FOR {each $d_j \in D$}
\STATE /* By instruction name and parameters */
\STATE \textit{sort}($d_j$)
\ENDFOR
\end{algorithmic}
\end{algorithm}

\begin{algorithm}[h!]
\caption{Dependency Detection (hasDependency())}\label{alg:dependency_determination}
\begin{algorithmic}
\REQUIRE Two logical instructions, $l_0$ and $l_1$
\STATE /* Determine qubits used in both operations */
\STATE $Q_0 \leftarrow \text{\textit{getQubits}}(l_0)$
\STATE $Q_1 \leftarrow \text{\textit{getQubits}}(l_1)$
\FOR{each $q_i \in Q_0$}
\FOR{each $q_j \in Q_1$}
\IF {$q_i == q_j$}
\STATE return \textbf{True}
\ENDIF
\ENDFOR
\ENDFOR
\STATE return \textbf{False}
\end{algorithmic}
\end{algorithm}

\subsection{Decomposition Pass} \label{subsec:Decomposition_pass}
% Up to this point, all the instructions processed are still specified on a logical level. However, operations on quantum hardware can only execute operations on physcial qubits. Therefore, the logical instructions needs to be decomposed into physical instructions.

% There are two main approaches for this decomposition:

To execute operations on quantum hardware, we need to translate logical instructions into physical instructions that can be performed on physical qubits. This conversion process, known as decomposition, ensures that the logical operations are compatible with the hardware's capabilities. 

There are two approaches to this decomposition:
\begin{enumerate}
    \item \textbf{Naive Decomposition:} This straightforward method uses a lookup table to map each logical instruction to its corresponding physical operations based on predefined rules.
    \item \textbf{Optimized Decomposition:} To decrease execution time, this approach leverages the hardware's parallel and pipelined capabilities. Our algorithm applies multiple strategies to optimize instruction sequences for better parallelism and pipelining.
\end{enumerate}

The Naive decomposition can be executed by using the logical to physical mapping presented in Sec. \ref{subsubsec:baseline_decomp}. However, many qubit operations can be optimized beyond the basic mapping to improve parallel execution and pipelining, by using the techniques described in Sec. \ref{subsubsec:optimized_CX_decomp}, Sec. \ref{subsubsec:optimized_parallel_sequence_decomp}, and Sec. \ref{subsubsec:optimized_pipelined_sequence_decomp}. In order to know which of these techniques can be used on a set of instructions, the compiler needs to know which instructions are in parallel, pipelined or sequential (Section \ref{subsec:parallel_detect}). The compiler also needs to know the parallel sequence length for the instructions that are in parallel (Section \ref{Subsec:parallel_seq_length}). At last the compiler decides which decomposition strategy to execute (Section \ref{subsec:decomposition_execution})
% The choice of decomposition approach is influenced by the distribution mode and the encoding scheme employed by the hardware.

% We utilize three optimization strategies:
% \begin{enumerate}
%     % \item \textbf{Improved Single Instruction Decomposition:} : Even if an instruction stands alone and isn't a special case, it can be decomposed more efficiently whenever possible.
%     \item \textbf{Leveraging Parallel Sequences:} By decomposing sequences of logical instructions in a pairwise parallel manner, we can achieve greater parallelism in the resulting physical instructions (see Sec. \ref{subsubsec:optimized_parallel_sequence_decomp}).
%     \item \textbf{Pipeline Potential:} We take advantage of the potential to pipeline two or more parallel sequences (see Sec. \ref{subsubsec:optimized_pipelined_sequence_decomp}).
%     \item \textbf{Enhanced CX Gate Decomposition:} Within the decomposition of a CX gate, we increase the number of parallel or pipelined instructions by doing pairwise parallel or pipelined decomposition. We can also use the physical \rx{} and \ry{} gates in the decomposition of a CX gate to pipeline or parallelize (see Sec. \ref{subsubsec:optimized_CX_decomp}).

% \end{enumerate}
% To choose the appropriate decomposition method, the compiler must know if instructions are in parallel, pipelined or serial. It also needs to know the lenghts of the parallel instructions sets 

\subsubsection{Detect parallel, pipelined, and serial sequences}  \label{subsec:parallel_detect} %\hfill \\
To choose the appropriate decomposition method, the compiler must determine whether instructions should be executed in parallel, pipelined, or serially. This is achieved by analyzing their dependencies.
The compiler checks if two instructions can be parallelized by verifying they have the same opcode and parameters and do not target the same qubits. Instructions are deemed pipelined when they do not have the same opcode, but act on different qubits and instructions are deemed sequential when they act on the same qubits.
% are not It evaluates if the instructions are pipelined by ensuring they do not operate on the same qubits. It evaluates if the instruction is sequential when two following instructions act on the same qubit.
% If they cannot be parallelized, it evaluates if they can be pipelined by ensuring they do not operate on the same qubits. If neither condition is met, the instructions are scheduled to execute serially.

\subsubsection{Parallel Sequence Lengths} \label{Subsec:parallel_seq_length}%\hfill \\
In Sec. \ref{subsubsec:optimized_pipelined_sequence_decomp} and Sec. \ref{subsubsec:optimized_parallel_sequence_decomp} decomposition techniques are presented which make use of parallel sequences. The compiler needs to know the length of these parallel sequences in order to use the decomposition techniques. The compiler finds parallel sequences by first
% The compiler needs to be capable of identifying parallel sequences, because they are used to apply a specific decomposition technique.  It starts by
finding two instructions that can be parallelized, then expands this set by including additional instructions that meet parallelization criteria, forming a complete parallel sequence. At this stage, the compiler focuses on maximizing parallel execution without considering the specific constraints introduced by different encoding schemes. The process for detecting parallel sequences is detailed in Alg. \ref{alg:parallel_sequence_detector}.

\begin{algorithm}[h!]
\caption{Parallel Sequence Detection}\label{alg:parallel_sequence_detector}
\begin{algorithmic}
\REQUIRE The list of logical instructions, $L$
% \STATE Determine the list of logical instructions: \textbf{instr\_logical}
\STATE /* Create empty qubit list */
\STATE $Q \leftarrow \{ \}$
% \STATE create empty qubit list: \textbf{qubit\_list}
\FOR{each $l_i \in L$}
% \FOR {Every instruction in instr\_logical}
\IF {\textit{isParallel}($l_i, l_{i+1}$)}
% \IF {Current instruction parallel with next instruction}
\STATE /* Get current instruction's qubit(s) */
\STATE $q \leftarrow \text{\textit{getQubits}}(l_i)$
\IF {$q \notin Q$}
\STATE \textit{append}($q, Q$)
% \STATE Store current qubit in \textbf{qubit\_list}
\ELSE 
\STATE /* End the current parallel sequence */
\STATE \textbf{break}
\ENDIF
\ENDIF
\ENDFOR
\STATE \textbf{Return} $L$
\end{algorithmic}
\end{algorithm}

\subsubsection{Decomposition Execution} \label{subsec:decomposition_execution}%\hfill\\ 
In Sec. \ref{sec:logical-gate-operations}, we explored various decomposition methods. We now shift our focus to identifying situations where improved decompositions can be performed. The compiler iterates over each logical instruction and decides which decomposition strategy to apply based on the following criteria:

We utilize three optimization strategies:
\begin{enumerate}
    % \item \textbf{Improved Single Instruction Decomposition:} : Even if an instruction stands alone and isn't a special case, it can be decomposed more efficiently whenever possible.
    \item \textbf{Leveraging Parallel Sequences:} By decomposing sequences of logical instructions in a pairwise parallel manner, we can achieve greater parallelism in the resulting physical instructions (see Sec. \ref{subsubsec:optimized_parallel_sequence_decomp}).
    \item \textbf{Pipeline Potential:} We take advantage of the potential to pipeline two or more parallel sequences (see Sec. \ref{subsubsec:optimized_pipelined_sequence_decomp}).
    \item \textbf{Enhanced CX Gate Decomposition:} Within the decomposition of a CX gate, we increase the number of parallel or pipelined instructions by doing pairwise parallel or pipelined decomposition. We can also use the physical \rx{} and \ry{} gates in the decomposition of a CX gate to pipeline or parallelize (see Sec. \ref{subsubsec:optimized_CX_decomp}).

\end{enumerate}

\begin{enumerate}
    \item \textbf{Sequential Instruction:} If an instruction is to be executed sequentially with the following instruction, the compiler applies the basic decomposition method, as outlined in Sec. \ref{subsubsec:baseline_decomp}. 
    \item \textbf{Parallel Sequence:} If an instruction is to be executed in parallel with the following instruction, it indicates the potential start of a parallel sequence. The compiler identifies the complete parallel sequence using the parallel sequence detection algorithm detailed in Alg. \ref{alg:parallel_sequence_detector}. After defining the length of a parallel sequence, the compiler can perform one of two optimized decompositions (Sec. \ref{subsubsec:optimized_pipelined_sequence_decomp}  Sec. \ref{subsubsec:optimized_parallel_sequence_decomp}) dependent on the set of instructions following the parallel sequence. If the following set of instructions is another parallel instruction set, the decomposition strategy described in Sec. \ref{subsubsec:optimized_parallel_sequence_decomp} can be used, otherwise the strategy described in Sec. \ref{subsubsec:optimized_pipelined_sequence_decomp} will be used.
    
    % the compiler needs to perform 2 different types of optimized decomposition, either it can decompose all the logical instructions into a set of parallel physical instructions, or it checks if the next set of instructions is also a set of parallel instructions and pipelines both sets of instructions.

    % Subsequently, the compiler checks if the execution of the parallel sequence can be pipelined with the upcoming instruction after the parallel sequence. If feasible, a second parallel sequence is created for the pipelined instruction. If the two sequences do not share any qubits, their execution can be pipelined and the decomposition is handled as presented in Sec. \ref{subsubsec:optimized_pipelined_sequence_decomp}. However, if there are shared qubits between the sequences, the initial parallel sequence is decomposed based on the distribution mode, as presented in Sec. \ref{subsubsec:optimized_parallel_sequence_decomp}. 
    
    The complete process is outlined in Alg. \ref{alg:semi_distributed}.
    \item \textbf{Pipelineable Instruction:} When the execution of two instructions can be pipelined, the compiler performs a pairwise pipelined decomposition. For \rx{} and \ry{} gates, this decomposition follows the same method as the parallel pipelined sequence, as explained in Sec. \ref{subsubsec:optimized_pipelined_sequence_decomp}. However, if the instruction is a CX instruction followed by an \rx{} or \ry{} gate, the instruction is decomposed as presented in Sec. \ref{subsubsec:optimized_CX_decomp}.
    \item \textbf{Special subBIT\_ncBIT Case:} In the subBIT\_ncBIT encoding scheme, the hardware allows parallel execution both within and across subnets. 
    Due to the fact that every qubit is placed on or across a subnet and depencies can only happen when operations act on the same node, parallelism can be achieved between any two logical instructions that do not act on the same qubit. The compiler exploits this capability to enhance parallelism, as outlined in Alg. \ref{alg:subBit_NCBit}. 
\end{enumerate}

\begin{algorithm}[h!]
\caption{Decomposition Selection Algorithm}\label{alg:semi_distributed}
\begin{algorithmic}
% \STATE Determine the list of logical instructions: 
% \textbf{instr\_logical}
\REQUIRE The list of logical instructions, $L$

\FOR {each $l_i \in L$}
\IF {\textit{isSequential}($l_i, l_{i+1}$)}
\STATE \textit{basicDecomposition}($l_i$)
\ELSIF {\textit{inParallelSequence}($l_i$)}
\IF {\textit{isPipelined}($l_i, l_{i+1}$)}
\STATE \textit{parallelPipelinedDecomposition}($l_i, l_{i+1}$)
\ELSE
\STATE \textit{parallelSequenceDecomposition}($l_i, l_{i+1}$)
\ENDIF
\ELSIF {\textit{isPipelined}($l_i,l_{i+1}$)}
\IF {$l_i == \text{CX}_L$}
\IF {$l_{i+1} == R_X(\theta)$ or $l_{i+1} == R_Y(\theta)$}
\STATE \textit{pipelinedInclusionDecomposition}($l_i,l_{i+1}$)
\ENDIF
\ELSE
\STATE \textit{pipelinedDecomposition}($l_i, l_{i+1}$)
\ENDIF
\ENDIF
\ENDFOR
\end{algorithmic}
\end{algorithm}

% \begin{algorithm}[H]
% \caption{Bit Bit case}\label{alg:semi_distributed}
% \begin{algorithmic}
% \STATE Determine the list of logical instructions: \textbf{instr\_logical}
% \STATE Create empty subnet list
% \STATE Create empty qubit list
% \IF {current instruction in parallel with next instruction}
% \STATE Find maximum parallel sequence
% \FOR {Instructions in parallel sequence}
% \STATE Determine which subnet the instruction is working on
% \STATE Add subnet and qubits to list
% \ENDFOR
% \STATE Count the amount of times the different subnets have been used
% \STATE Order subnets with highest amount of usages first
% \FOR {subnets in ordered subnet list}
% \STATE Perform decomposition with current subnets and corresponding qubits
% \ENDFOR
% \ENDIF
% \end{algorithmic}
% \end{algorithm}
\begin{algorithm}[h!]
\caption{subBIT\_ncBIT: Parallel Decomposition}\label{alg:subBit_NCBit}
\begin{algorithmic}
% \STATE Determine the list of logical instructions: \textbf{instr\_logical}
\REQUIRE The list of logical instructions, $L$
\STATE /* Create empty subnet and qubit lists */
\STATE $S \leftarrow \{ \},  Q \leftarrow \{ \}$
\FOR{each $l_i \in L$}
    \IF{\textit{isParallel}($l_i, l_{i+1}$)}
        \STATE /* Find maximum length parallel sequence */
        \STATE $S_{\parallel} \leftarrow \text{\textit{maxParallelSeq}($l_i$)}$
        \FOR{each instruction $l_k \in S_{\parallel}$}
            \STATE /* Get target qubits and subnets */
            \STATE $q \leftarrow \text{\textit{getQubits}}(l_k)$
            \STATE $s \leftarrow \text{\textit{getSubnets}}(q)$
            % \STATE /* Add qubits and subnets to respective lists */
            \STATE \textit{append}($q, Q$)
            \STATE \textit{append}($s, S$)
        \ENDFOR
    \ENDIF
\ENDFOR
\STATE /* Count occurrences of each subnet in the list */
\STATE $C \leftarrow \text{\textit{getCounts}}(S)$
\STATE /* Sort subnet list by count */
\STATE \textit{sort}($S, C$)
\FOR{$s_i \in S$}
    \STATE getQubitsBySubnet(S)
    \STATE PerformDecomposition()
\ENDFOR
\end{algorithmic}
\end{algorithm}

% \subsubsection{Fully distributed}  \hfill \\
\subsection{Subnet Pass}  \label{subsec:Subnet_pass}
The subnet pass ensures that the final output instructions are appropriately marked for parallel execution based on the hardware configuration. This marking informs the hardware that these instructions can be executed simultaneously. The decision to parallelize instructions depends on the chosen encoding scheme and the size of the subnets.

% The subnet pass ensures that instructions generated as final outputs are marked for parallel execution based on the configuration of the hardware. This clear marking informs the underlying system that these instructions can be executed in parallel. The decision on whether instructions can be parallelized depends on encoding scheme and size of the subnets.
The subnet pass is specifically designed for the subID\_ncBIT and subBIT\_ncID encoding schemes, because the subBIT\_ncBIT case is already resolved during the decomposition pass. This is because the subBIT\_ncBIT case needs the instructions to be re-ordered based on the configuration of the hardware. In this pass the instructions are not re-ordered but only marked for the underlying hardware in order to execute the insturctions as intended.

% The decomposition pass already takes into account the addressing modes of the subBIT\_ncBIT case and therefore this pass does not add any value to this case.

The subnet pass operates by following two steps:
\begin{enumerate}
    % \item \textbf{Creating Subnets:} The subnet pass creates subnets based on the chosen encoding schemes. These subnets serve as a basis for maximizing parallelism.
    \item \textbf{Determining Parallel instructions:} In a parallel sequence, the pass identifies which subnet the qubits belong to by making use of Eq. \ref{eq:subnet_subnetID} and Eq. \ref{eq:subnet_subnetbit}, as described in Sec. \ref{sec:subnet-organization}. The qubits are then defined as parallel either by being in the same subnet or in different subnets, corresponding to the subID\_ncBIT and subBIT\_ncID encoding schemes, respectively.
    \item \textbf{Reduce Parallel Sequences based on configuration:} 
    If an instruction cannot be executed in parallel with preceding instructions due to hardware constraints, the current sequence of parallel instructions is ended, and a new sequence begins with the current instruction. This method optimizes the number of parallel instructions within a specified subnet size. Consequently, the length of each set of parallel instructions is maximized to match the hardware's capacity for parallel execution.
\end{enumerate}
\section{Run Time Analysis} \label{sec:runtime-analysis}
% Having defined the hardware design possibilities and the functionality needed in the compiler to transform quantum algorithms into a format compatible with parallel execution, we now present run time analyses for each instruction sequence type.

% a method for evaluating the throughput vs. single-instruction latency tradeoff of different addressing modes. For a particular quantum algorithm, this method determines the maximum amount of parallelization overhead, $\delta$, that can be tolerated for a given parallelizability factor, $\rho$, before the parallelized version of a quantum algorithm takes longer to execute than the default version. 

\begin{figure*}
    \centering
    \begin{subfigure}[b][][b]{0.35\textwidth}
        \includegraphics[width=\textwidth, keepaspectratio]{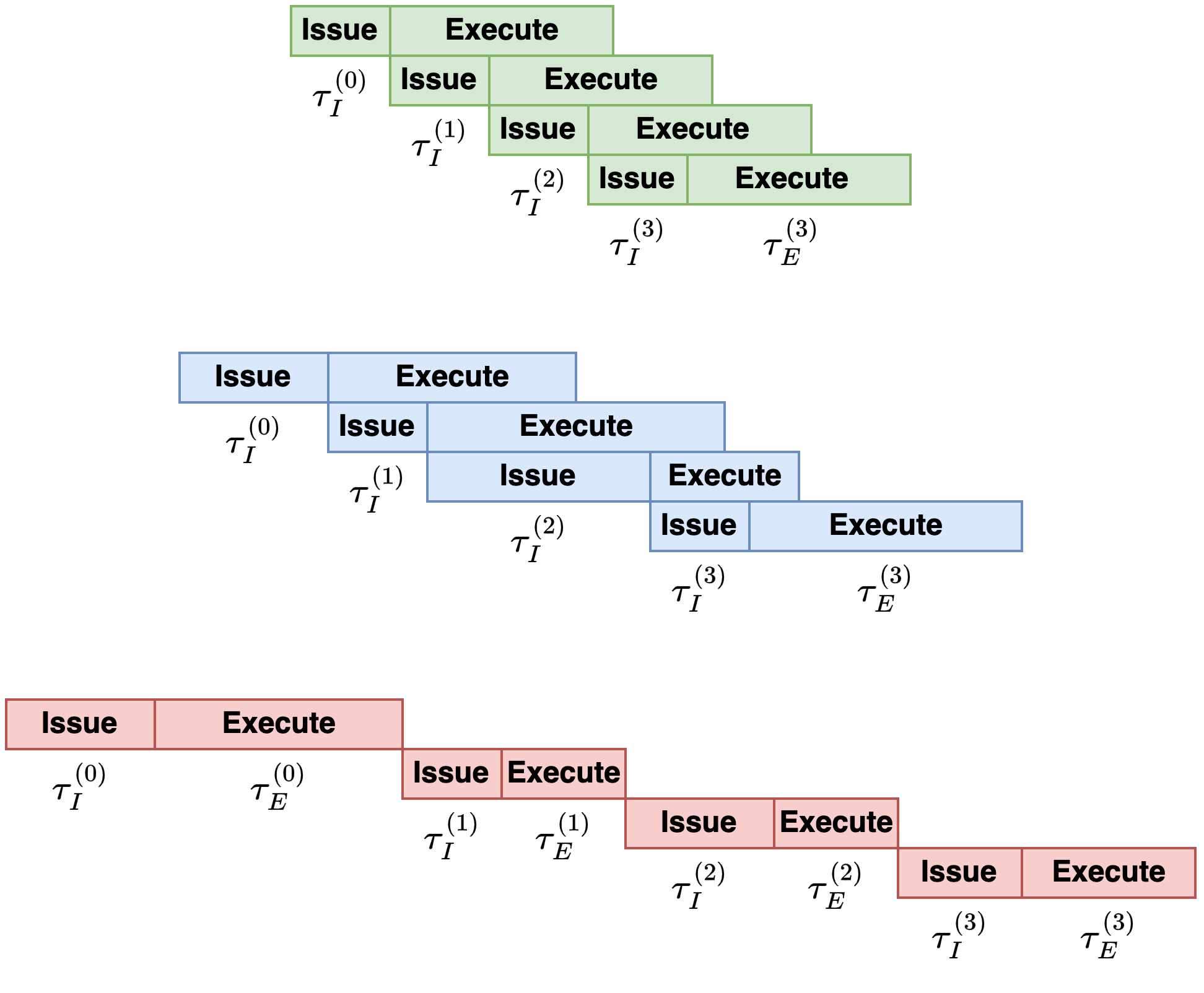}
        \caption{Before parallelizing execution.}
        \label{fig:instr-seqs-ser}
    \end{subfigure}
    \hspace{1cm}
    \begin{subfigure}[b][][b]{0.35\textwidth}
        \includegraphics[width=\textwidth, keepaspectratio]{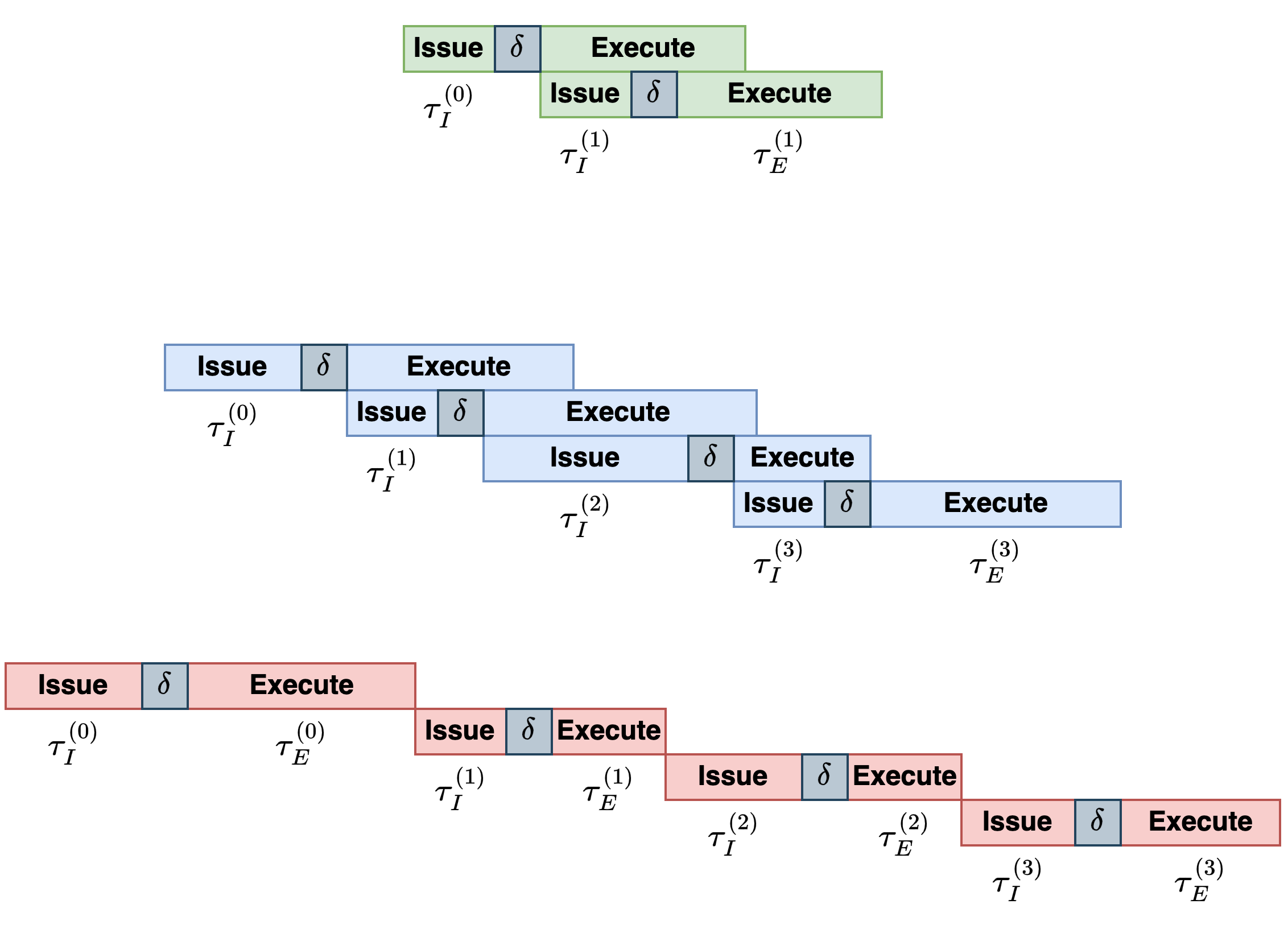}
        \caption{After parallelizing execution.}
        \label{fig:instr-seqs-par}
    \end{subfigure}
    \caption{Execution graphs for parallel (green), pipelined (blue), and serial (red) instruction sequences of length $N=4$. After adding hardware support for parallel execution, parallel instruction sequence lengths are reduced by a factor of $\rho$ ($\rho=2$ is assumed here). In all three instruction sequence types, individual instruction runtimes are increased by $\delta$ cycles due to longer address transmission.}
    \label{fig:instr-seqs}
    % \Description{This figure depicts the execution time of a parallel, pipelined, and serial instruction sequence before and after parallelizing instruction execution. After parallelizing instruction execution, the number of instructions in the parallel instruction sequence is reduced, while the single-instruction latency is increased for all instruction sequence types.}
\end{figure*}

Having defined the hardware design possibilities and the functionality needed in the compiler to transform quantum programs into a format compatible with parallel execution, we now present run time analyses for each instruction sequence type. Specifically, we will describe mathematically how parallelization of a quantum program affects the run time of parallel, pipelined, and serial instruction sequences, individually, as visualized in Fig. \ref{fig:instr-seqs}. This analysis will then be used in Sec. \ref{Sec:results} to model the run times of entire quantum programs such that a comparison can be made between the run times of the default and parallel versions of a program. To simplify our run time discussions, we divide the run time of a single instruction according to the two instruction phases identified in Sec. \ref{sec:system-dependencies}:

\begin{enumerate}
    \item \textbf{Issue Time}: the latency associated with the issue phase of an instruction using a SISD execution model. The issue time of an instruction is denoted $\tau_I$.
    \item \textbf{Execution Time}: the latency associated with the execution phase of an instruction. The execution time of an instruction is denoted $\tau_E$.
\end{enumerate}

\noindent
In this analysis, both $\tau_I$ and $\tau_E$ are measured in clock cycles. The issue and execution times for the $j$th instruction in a sequence or algorithm are denoted $\tau_I^{(j)}$ and $\tau_E^{(j)}$, respectively.

\subsection{Parallel Sequence Run Time}
Since, by definition, a parallel instruction sequence does not contain any node controller dependencies before instruction execution is parallelized, a parallel sequence executes as though it were a pipelined sequence. Therefore, to determine the total run time of a parallel sequence in the default version of a quantum program, the issue time of each instruction as well as the execution time of the last instruction in the sequence must be taken into account. This leads to the default run time expression for a parallel sequence of $N$ instructions, $T_{\parallel}^{def}$, shown in Eq. \ref{eq:runtime-par-seq-def}. The final simplified form is justified since every instruction in a parallel sequence performs the same operation with the same set of parameters. Therefore, the issue time of every instruction in the sequence is the same.

\begin{equation}
    T_{\parallel}^{def} = \Sum{j=0}{N-1} \tau_{I}^{(j)} + \tau_{E}^{(N-1)} = N\tau_{I}^{(0)} + \tau_{E}^{(N-1)} \label{eq:runtime-par-seq-def}
\end{equation}

\noindent
For the parallelized version of a quantum program, the length of the parallel sequence will be reduced by a factor of $\rho$. In general, the new length of the sequence is $\left \lceil \frac{N}{\rho} \right \rceil$ instructions. Simultaneously, the issue time of each instruction increases by $\delta$ cycles. The run time for the parallelized version of the sequence, $T_{\parallel}^{par}$, is shown in Eq. \ref{eq:runtime-par-seq-par}, where $N' = \left \lceil \frac{N}{\rho} \right \rceil$.

\begin{equation}
    T_{\parallel}^{par} = \Sum{j=0}{N' - 1} \left ( \tau_{I}^{(j)} + \delta \right ) + \tau_{E}^{(N'-1)} = N' \left ( \tau_{I}^{(0)} + \delta \right ) + \tau_{E}^{(N'-1)} \label{eq:runtime-par-seq-par}
\end{equation}

\subsection{Pipelined Sequence Run Time}
For the case of a pipelined sequence of instructions, we must again account for the issue time of each instruction and the execution time of the final instruction. In the default version of a quantum program, the run time of the pipelined sequence, $T_p^{def}$, takes the form shown in Eq. \ref{eq:runtime-pipe-seq-def}. In this case, the final expression cannot be further simplified, since there is no guarantee that all instructions in a pipelined sequence have the same parameters. Therefore, their issue times may differ. In reality, it is also not guaranteed that the final instruction in the sequence is the last to finish execution, so the total run time may be extended by some number of cycles. However, particularly as $N \rightarrow \infty$, these added cycles contribute negligibly to the total run time, so they can be ignored.

\begin{equation}
    T_p^{def} = \Sum{j=0}{N-1} \tau_I^{(j)} + \tau_E^{(N-1)} \label{eq:runtime-pipe-seq-def}
\end{equation}

In the parallelized version of a quantum program, the number of instructions in a pipelined sequence does not change. On
the other hand, since the address encoding in the instructions has changed, the issue time of each instruction is still extended due to the parallelism overhead, $\delta$. These effects are captured in Eq. \ref{eq:runtime-pipe-seq-par}.

\begin{equation}
    T_p^{par} = \Sum{j=0}{N-1} \left ( \tau_{I}^{(j)} + \delta \right ) + \tau_{E}^{(N-1)} = N\delta + \Sum{j=0}{N-1} \tau_{I}^{(j)} + \tau_{E}^{(N-1)} \label{eq:runtime-pipe-seq-par}
\end{equation}

\subsection{Serial Sequence Run Time}
When considering serial sequences of instructions, we must account for both the issue and execution time of every instruction in the sequence. This leads to the run time of a serial sequence in the default version of a quantum program, $T_s^{def}$, which is shown in Eq. \ref{eq:runtime-ser-seq-def}.

\begin{equation}
    T_s^{def} = \Sum{j=0}{N-1} \left( \tau_I^{(j)} + \tau_E^{(j)} \right) \label{eq:runtime-ser-seq-def}
\end{equation}

As was the case with pipelined sequences, the number of instructions in a serial sequence does not change in the parallelized version of a quantum program, and each instruction's issue time is extended by $\delta$ cycles. This leads to the run time expression for a serial sequence in the parallelized version of a program, $T_s^{par}$, shown in Eq. \ref{eq:runtime-ser-seq-par}.

\begin{equation}
    T_s^{par} = \Sum{j=0}{N-1} \left( \tau_I^{(j)} + \delta + \tau_E^{(j)} \right) = N\delta + \Sum{j=0}{N-1} \left( \tau_I^{(j)} + \tau_E^{(j)} \right) \label{eq:runtime-ser-seq-par}
\end{equation}

\section{Results and Discussion} \label{Sec:results}
In this section, we evaluate how well the hardware and compiler designs presented in Sec. \ref{Sec:Hardware_design} and Sec. \ref{sec:Compiler_design} are able to improve the performance of a set of benchmark quantum programs. For this purpose, we model a quantum program as a collection of parallel, pipelined, and serial sequences of instructions and accumulate the run times of each sequence according to the expressions derived in Sec. \ref{sec:runtime-analysis}.

% We also evaluate how well the parallelism boundary condition presented in Section \ref{sec:Evaluation_tool} is able to identify the optimal design candidate for the set of benchmarks.

\subsection{Methodology}
% Here we talk about how we determined:
% \begin{enumerate}
%     \item our system setup + how that determines where we get the issue/execution times from
%     \item the addressing modes we use
%     \item the benchmarks we use
% \end{enumerate}

\subsubsection{System Setup}
For our analysis, we target the compiler and hardware for a distributed quantum computing system using nitrogen vacancy (NV) centers in diamond \cite{Diamond}. Each node in the system's network corresponds to a single NV center. In the semi-distributed mode, we assume two Carbon-13 nuclei in the vicinity of each NV center are used as data qubits, while in the fully distributed case, we assume only one Carbon-13 nucleus is used as a data qubit. In both cases, the trapped electron in each NV center is used as an ancilla qubit. 

In order to measure the performance of the hardware design and compiler, we developed a Python script which utilizes the sequence run time expressions developed in Sec. \ref{sec:runtime-analysis} to calculate the total run time of entire quantum programs. As a baseline, the run time of a quantum program is calculated without compiler optimizations using a SISD execution model. Three different categories of speedup are then calculated.

\begin{itemize}
    \item \textbf{Compiler Speedup}: measured by comparing the run time of the baseline model to the run time after applying compiler pipelining optimizations but before enabling parallel execution in the hardware. 
    \item \textbf{Hardware Speedup:} measured by comparing the run time of the programs after compiler pipelining optimizations have been performed to the run time when combining compiler optimizations and parallel execution support in the hardware.
    \item \textbf{Combined Speedup}: measured by comparing the run time of the baseline model to the run time when combining compiler optimizations and parallel execution support in the hardware.
\end{itemize}

% Subsequently, the \textit{compiler speedup} is measured by calculating the run time after applying compiler pipelining optimizations but before enabling parallel execution in the hardware. Finally, the run time is calculated when combining compiler optimizations and parallel execution support in hardware. We measure the additional \textit{hardware speedup} by calculating the speedup of this run time relative to the isolated compiler run time. Furthermore, we obtain a \textit{combined speedup} by calculating the speedup of this run time relative to the baseline model.

For the classical hardware, we assume that the central and node controllers operate at a frequency of 10 MHz. The central controller issues instructions to the node controller network via a network interface of width $L = 16$ bits. Verilog implementations of the node controller and signal generators have been developed as part of this research. The node controller utilizes a three-stage pipeline for decoding and configuring execution of the quantum instructions. To perform operations on the Carbon-13 and electron qubits, respectively, the node controller interfaces with separate radio-frequency (RF) and microwave (MW) generators. These RF and MW signal generators operate with constant delays of 1.32 $\mu$s and 396 ns, respectively. With this information about the system, we can set the total run time of each instruction in a quantum algorithm by calculating their total issue and execution times; these values are presented in Tab. \ref{tab:issue-execution-times}. 

We now briefly explain derivations for these values. To determine the issue times for each instruction, a constant two-cycle issue delay is assumed for transmitting the instruction opcode and address in the SISD execution model. An additional cycle of delay is introduced by each 16-bit parameter required by a given instruction (e.g., signal phase or angle of rotation). The \rx{}, \ry{}, and $\text{CR}_X(\theta)$ gate execution times are calculated by determining the pipeline delay through the node controller and the cycles required by the MW and RF generators for generating signals. Furthermore, we assume a virtual $Z$ gate implementation \cite{PhysRevA.96.022330} when calculating the $R_Z(\theta)$ gate execution time. For the CX gate, its issue/execution time is computed as the sum of the issue/execution times of each instruction in the sequence from Fig. \ref{fig:full_cx_decomp}. Finally, since fixed protocols for the entangle and measure instructions are not yet finalized for diamond NV centers, we derive reasonable estimations for the execution times of these instructions from the literature. For the measure instruction, a value of 40 $\mu$s, or 400 clock cycles, was found in \cite{robledo2011high}. For the entangle instruction, since a variable number of attempts is required before entanglement is successfully heralded, an execution time value can be derived using the formula $t_{ent} = t_{attempt} \times N$, where $t_{attempt}$ is the time it takes to perform a single entanglement attempt and $N$ is the average number of attempts needed. In \cite{pompili2021realization}, $t_{attempt}$ was found to be 5.8 $\mu$s, while we choose $N = 20$ as an evaluation for future diamond NV center quantum systems. This gives an execution time of 11.6 $\mu$s, or 1160 clock cycles, for the entangle instruction.
% Where the time it takes for a single attempt is equal to $0.0058ms$ (derived from \cite{pompili2021realization}) N is the number of entanglement attempts

\begin{table}[h!]
\centering
\begin{tabular}{|>{\centering\arraybackslash}m{0.213\linewidth}|>{\centering\arraybackslash}m{0.25\linewidth}|>{\centering\arraybackslash}m{0.35\linewidth}|} 
\hline
\textbf{Instruction} & \textbf{Issue Time (cycles)} & \textbf{Execution Time (cycles)}  \\ 
\hline
$R_X(\theta)$          & 5               & 62                 \\ 
\hline
$R_Y(\theta)$          & 5               & 62                 \\ 
\hline
$R_Z(\theta)$          & 3               & 11                  \\ 
\hline
$\text{CR}_X(\theta)$         & 4               & 62                 \\ %1206
\hline
entangle    & 3               & 1160                \\ 
\hline
measure     & 2               & 400                \\ 
\hline
\end{tabular}
\caption{Issue and execution times for each instruction type.}
\label{tab:issue-execution-times}
\end{table}

\subsubsection{Addressing Mode Selection}
Using the system described above, we consider support for 100-500 logical qubits, since it corresponds to a system of 400-2000 physical qubits; this extends the analysis well into the range of system sizes projected for intermediate-term quantum computing systems. This number of physical qubits requires a network of 1000 node controllers in a semi-distributed mode and a network of 2000 node controllers in a fully distributed mode. To simplify our analysis, we use a node controller network of 1024 node controllers in the semi-distributed mode and a network of size 2048 node controllers in the fully distributed mode.

Several addressing modes are possible within these network sizes depending on the number of subnets used to partition the node controller network. Within the subID\_ncBIT and subBIT\_ncBIT address encoding schemes, we increment the number of subnets in each addressing mode by powers of 2 in the range from 1 up to 512 for the semi-distributed mode and from 1 up to 1024 for the fully distributed mode. Within the subBIT\_ncID address encoding scheme, we vary the number of subnets by powers of 2 in the range from 2 to 1024 for the semi-distributed mode and from 4 to 2048 in the fully distributed mode. This ensures that there is a minimal parallelizability factor of $\rho = 2$ in every design considered.

\subsubsection{Benchmark Selection}
The QASMBench \cite{li2022qasmbench} and MQTBench \cite{quetschlich2023mqtbench} benchmark suites are used to evaluate the hardware architecture and compiler designs. Both suites contain quantum programs of varying size which are expected to be relevant for future quantum computing systems. Moreover, MQTBench also features synthetically constructed quantum programs which allow for analyzing different instruction sequence patterns. The 18 benchmarks listed in Tab. \ref{tab:benchmarks} were selected since the numbers of logical qubits they use are within the range of interest. Tab. \ref{tab:benchmarks} contains only a brief description of each benchmark, more detailed information can be found in \cite{li2022qasmbench} and \cite{quetschlich2023mqtbench}.

\begin{table*}[ht!]
    \centering
    \small
    \begin{tabular}{|>{\centering\arraybackslash}m{0.15\linewidth}|>{\centering\arraybackslash}m{0.2\linewidth}|>{\centering\arraybackslash}m{0.55\linewidth}|} 
        \hline
        \textbf{Benchmark} & \textbf{Number of  Logical Qubits} & \textbf{Description} \\ \hline
        \textbf{\texttt{adder}} & 433 & Quantum Ripple-Carry Adder \cite{li2022qasmbench}. \\ \hline
        \textbf{\texttt{bv}} & 280 & Execution of the Bernstein-Vazirani Algorithm \cite{li2022qasmbench}. \\ \hline
        \textbf{\texttt{cat}} & 260 & Creating a superposition of two states with opposite phase \cite{li2022qasmbench}. \\ \hline
        \textbf{\texttt{ghz}} & 255 & Creating a maximally entangled Greenberger-Horne-Zeilinger state \cite{li2022qasmbench}. \\ \hline
        \textbf{\texttt{ising}} &  420 & Simulation of an Ising model \cite{li2022qasmbench}. \\ \hline
        \textbf{\texttt{qugan}} & 395 & Quantum generative adversarial network \cite{li2022qasmbench}. \\ \hline
        \textbf{\texttt{ae}} & 130 & Amplitude estimation for a certain quantum state \cite{quetschlich2023mqtbench}. \\ \hline
        \textbf{\texttt{dj}} & 130 & Execution of the Deutsch-Jozsa algorithm \cite{quetschlich2023mqtbench}. \\ \hline
        \textbf{\texttt{graphstate}} & 130 & Quantum circuit representation of graphs \cite{quetschlich2023mqtbench}. \\ \hline
        \textbf{\texttt{qnn}} & 130 & Quantum neural network \cite{quetschlich2023mqtbench}. \\ \hline
        \textbf{\texttt{qpeexact}} & 130 & Quantum phase estimation (exact) \cite{quetschlich2023mqtbench}. \\ \hline
        \textbf{\texttt{qpeinexact}} & 130 & Quantum phase estimation (inexact) \cite{quetschlich2023mqtbench}. \\ \hline
        \textbf{\texttt{random}} & 130 & Randomized quantum circuit \cite{quetschlich2023mqtbench}. \\ \hline
        \textbf{\texttt{realamprandom}} & 130 & Variational quantum eigensolver using a real amplitudes ansatz \cite{quetschlich2023mqtbench}. \\ \hline
        \textbf{\texttt{su2random}} & 130 & Variational quantum eigensolver using an SU2 ansatz \cite{quetschlich2023mqtbench}. \\ \hline
        \textbf{\texttt{swap\_test}} & 361 & Swap test routine used in quantum machine learning \cite{li2022qasmbench}. \\ \hline
        \textbf{\texttt{twolocalrandom}} & 130 & Variational quantum eigensolver using a Two Local ansatz \cite{quetschlich2023mqtbench}. \\ \hline
        \textbf{\texttt{wstate}} & 380 & Creation of an entangled W state \cite{li2022qasmbench}. \\ \hline  
    \end{tabular}
    \caption{Benchmarks used to evaluated the addressing mode options.}
    \label{tab:benchmarks}
\end{table*}
% A relevant set of benchmarks is required for determining the performance of the quantum computing system. To this end, the QASMBench \cite{li2022qasmbench} and MQTBench \cite{quetschlich2023mqtbench} benchmark suites are used. Both suites contain many quantum programs of varying size which are expected to be relevant for future quantum computing systems, while MQTBench also features synthetically constructed quantum programs which allow for analyzing different instruction sequence patterns. The 18 benchmarks listed in Table \ref{tab:benchmarks} were selected since the numbers of logical qubits they use are within the range of interest. Table \ref{tab:benchmarks} contains only a brief description of each benchmark, but more detailed information can be found in \cite{li2022qasmbench} and \cite{quetschlich2023mqtbench}.

\subsection{Compiler Speedup}\label{sec:Compiler_speedup}
\begin{figure}[ht!]
    \centering
    \begin{subfigure}{0.5\columnwidth}
    % \centering
        \includegraphics[width=\textwidth, keepaspectratio]{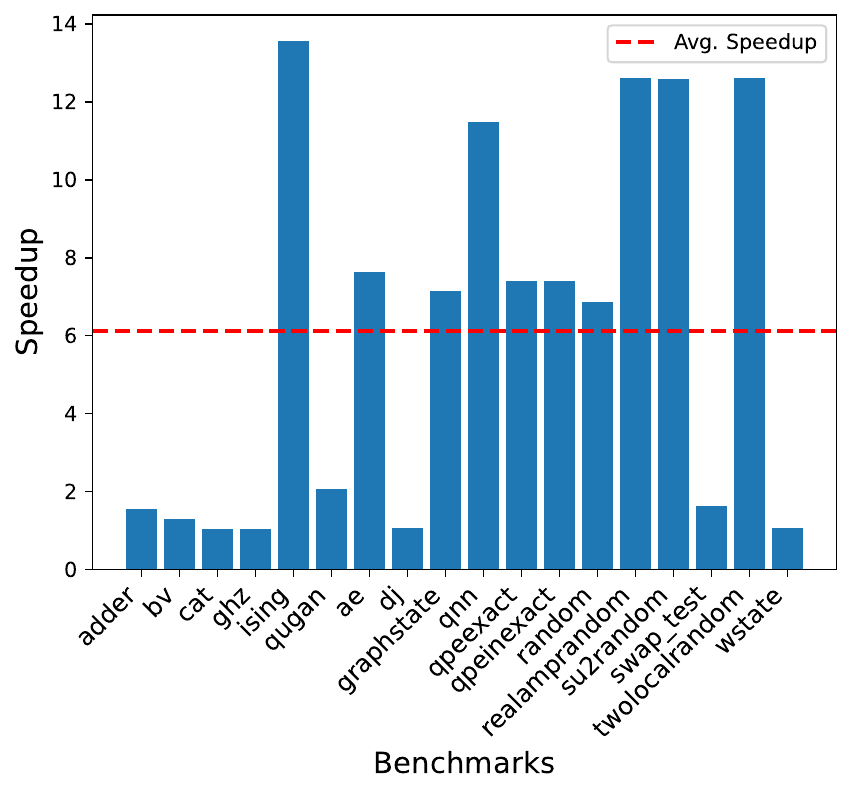}
        \caption{Semi-distributed mode.}
        \label{fig:compiler-semi-speedup}
    \end{subfigure}%
    \begin{subfigure}{0.5\columnwidth}
        % \hspace*{-3.2cm}
        \includegraphics[width=\textwidth, keepaspectratio]{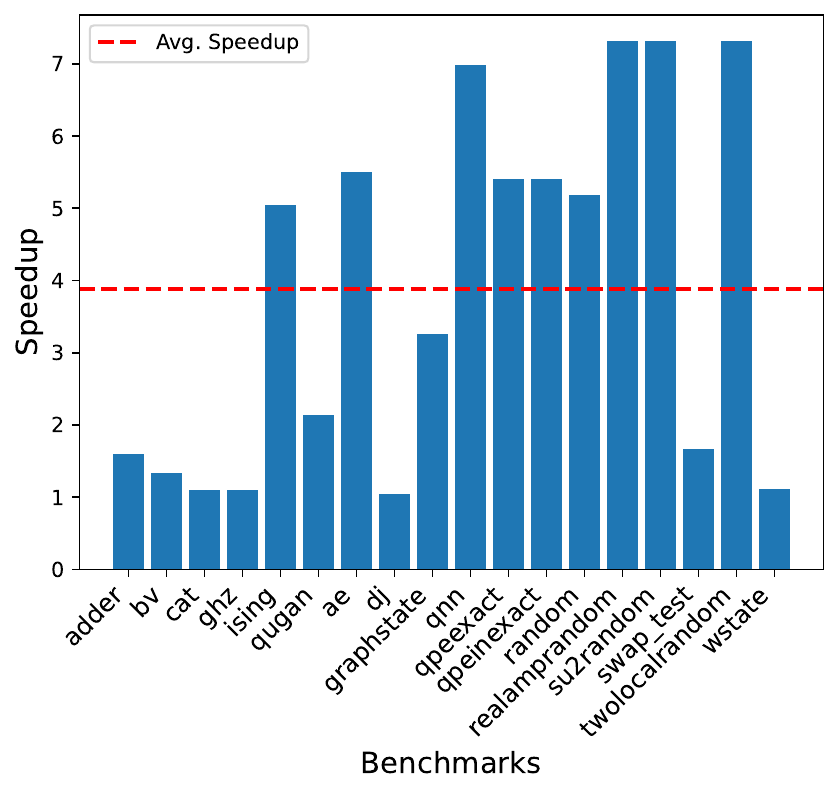}
        \caption{Fully distributed mode.}
        \label{fig:compiler-fully-speedup}
    \end{subfigure}%
    
    \caption{The speedup achieved for each benchmark using compiler optimizations. The red line corresponds to average speedup achieved across all benchmarks for the given distribution mode. The average speedup is 6.11x and 3.88x for the semi- and fully distributed mode respectively. A speedup value of 1 corresponds to no speedup observed for the benchmark.}
    \label{fig:total_speedup_compiler}
    % \Description{This figure shows the speedup achieved by applying compiler optimizations to each benchmark as well as the average speedup achieved across all benchmarks. Results are shown separately for executing in a semi-distributed mode versus a fully-distributed mode. The average speedup obtained is higher when executing in a fully-distributed mode than when executing in a semi-distributed mode.}
\end{figure}

The speedup results for the compiler optimizations in the semi- and fully distributed modes are presented in Fig. \ref{fig:compiler-semi-speedup} and \ref{fig:compiler-fully-speedup}, respectively. The results shows that, for every benchmark, a speedup was obtained. On average, the compiler achieves a speedup of 6.11x in the semi-distributed mode and 3.88x in the fully distributed mode. Furthermore, minimum and maximum speedup values of 1.03x and 13.55x, respectively, were observed for the semi-distributed mode, while minimum and maximum speedup values of 1.04x and 7.31x, respectively, were achieved for the fully distributed mode.
% The higher minimum speedup value for the fully distributed mode is expected because of the high number of CX gates in the program with minimal speedup and the manner in which CX gates are decomposed in the semi- and fully distributed cases.
The lower-performing programs usually contain more CX gates used for entangled state creation. These CX gates are decomposed in a more efficient manner for the fully distributed case since the four intermediate CX gates can be executed in parallel, resulting in four parallel sequences. In the semi-distributed case, only two intermediate CX gates can be executed in parallel.

\begin{table}[]
    \centering
     \begin{tabular}{|>{\centering\arraybackslash}m{0.2\linewidth}|>{\centering\arraybackslash}m{0.2\linewidth}|>{\centering\arraybackslash}m{0.2\linewidth}|>{\centering\arraybackslash}m{0.2\linewidth}|} \hline
       Encoding Scheme  & Minimum Speedup (semi,fully) & Maximum Speedup (semi,fully) & Peak Average Speedup (semi,fully) \\ \hline
       % subID\_ncBIT & (0.55,0.35) & (4.1,6.0) & (1.5,1.7) \\ \hline
       % subBIT\_ncID & (0.55,0.35) & (4.1,6.0) & (1.5,1.7)\\ \hline
       % subBIT\_ncBIT & (0.55,0.35) & (4.1,5.7) & (1.2,1.3) \\  \hline
       subID\_ncBIT & (0.55,0.35) & (4.1,6.0) & (1.96,2.48) \\ \hline
       subBIT\_ncID & (0.55,0.35) & (4.1,6.0) & (1.96,2.48)\\ \hline
       subBIT\_ncBIT & (0.55,0.35) & (4.1,5.7) & (1.55,1.90) \\  \hline
    \end{tabular}
    \caption{Minimum, maximum, and average hardware speedup values for the encoding schemes.}
    \label{tab:hw_speedup}
\end{table}

\subsection{Hardware Speedup}\label{sec:HW_speedup}
\begin{figure*}[ht!]
    \centering
    \includegraphics[width = 0.95\textwidth]{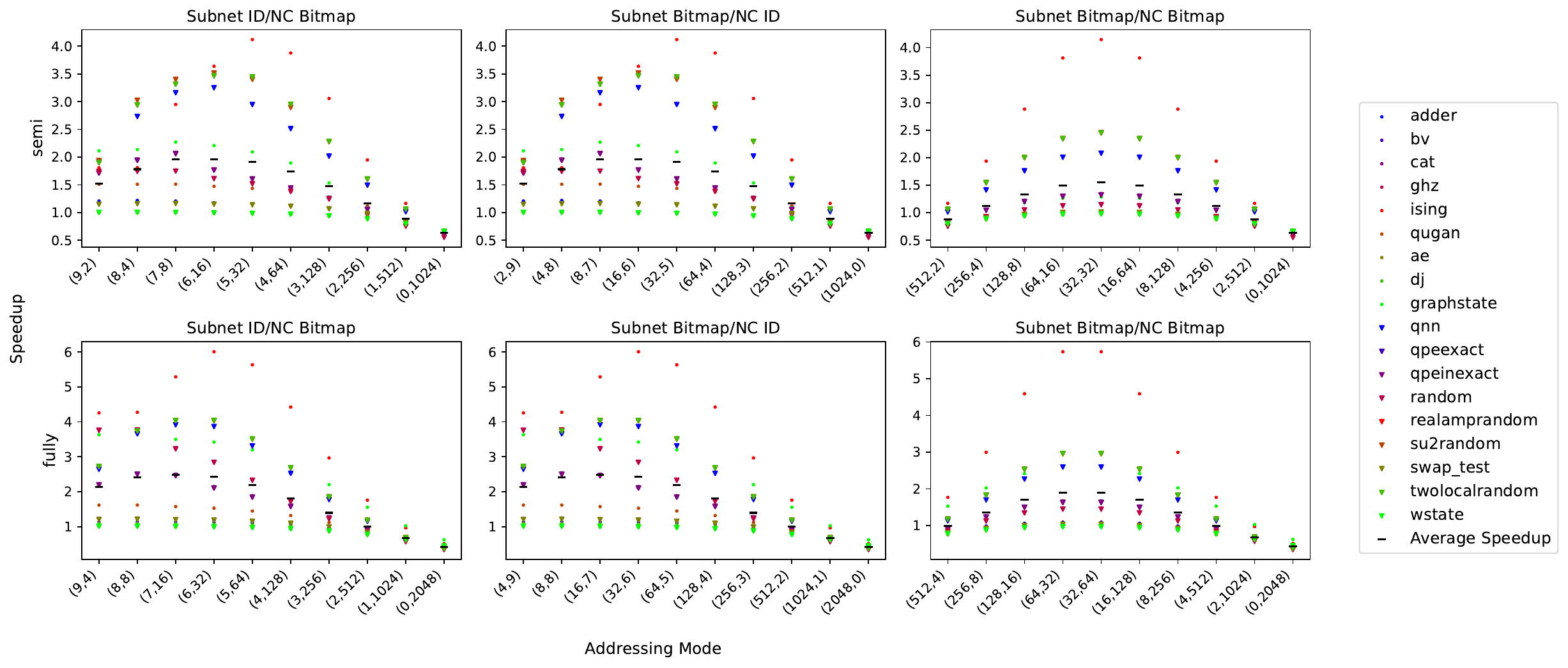}
    \caption{The additional speedup provided by the hardware design relative to the compiler. Each tick of the x-axis corresponds to a particular addressing mode, labeled $(W_S, W_{NC})$, where $W_S$ and $W_{NC}$ are the subnet and node controller address widths, respectively. For each addressing mode, colored data points mark the speedup value for a particular benchmark, while the black data point indicates the average speedup. The top row shows speedup values in the semi-distributed mode, while the bottom row shows speedup values in the fully distributed mode.}
    \label{fig:total_speedup_HW}
    % \Description{The plots shown in this figure show the additional speedup achieved by the proposed hardware design for each benchmark, as well as the average speedup across all benchmarks, for every combination of encoding scheme and distribution mode. The trend shows that, as the combined width of the subnet and node controller addresses initially increases, the achieved speedup also increases due to increased parallel execution. However, there is a clear optimal address width beyond which the speedup reduces again due to increased single-instruction latency.}
\end{figure*}

The hardware speedup is determined by comparing the total runtime of the system when the compiler and hardware optimization are both used versus when only the compiler is used. The speedup results for all addressing modes across each encoding scheme and distribution mode are presented in Fig. \ref{fig:total_speedup_HW}. For each addressing mode, the black marker in the plot represents the average speedup achieved across the benchmarks. The x-axis markers are labeled $(W_S, W_{NC})$ according to the subnet address width, $W_S$, and node controller address width, $W_{NC}$, of the given addressing mode. The minimum, maximum, and peak average speedup values obtained for each encoding scheme and distribution mode are presented in Tab. \ref{tab:hw_speedup}.

% The subID\_ncBIT encoding scheme results present a minimum speedup of 18 (3) and a maximum speedup of 28 (4) for the semi- (fully) distributed modes.

% The subBIT\_ncID encoding scheme results present a minimum speedup of 18 (3) and a maximum speedup of 28 (4) for the semi- (fully) distributed modes. 

% The subBIT\_ncBIT encoding scheme results present a minimum speedup of 0.8 (0.8) and a maximum speedup of 19 (3.5) for the semi- (fully) distributed modes. 

These results show that the hardware is capable of achieving speedup for every benchmark and encoding scheme if the addressing mode is chosen correctly. However, Fig. \ref{fig:total_speedup_HW} also shows a large discrepancy between the speedup obtained for certain sets of quantum programs. This discrepancy can be explained by the type of program that is being executed. Programs such as GHZ state creation are purely sequential in nature due to large dependency chains created by CX gates. In particular, we find that the target qubit in a CX gate becomes the control qubit in the next instruction which severely limits possible optimizations. We see, however, that programs such as the Ising model are highly parallelizable due to the low amount of dependencies between the instructions of the algorithm. 
Finally there are also other programs, such as amplitude estimation (marked \emph{ae}) and graph state, that have some parallel and some sequential parts, resulting in middle-of-the-road parallelizability.

From the results, we also observe the tradeoff discussed in Sec. \ref{Sec:Hardware_design} between instruction transmission overhead and parallel execution. Initially, as the bitmap-encoded portion of the instruction address is allowed to grow, there is a large increase in speedup. However, past an optimal address width, the speedup begins to decrease again as the penalty of instruction transmission overhead starts to outweigh the benefits of parallel execution. The optimal address width is based around the most common parallel sequence length.

We also note that the fully distributed case (bottom row of Fig. \ref{fig:total_speedup_HW}) achieves about twice the amount of speedup for programs with many sequential CX gates, again due to the different CX gate decompositions in the two distribution modes. Since each CX gate is comprised of a long series of instructions, the execution time of many CX gates will tend to dominate the total execution time, so it is unsurprising that the distribution mode has such a significant effect. Finally, we observe that the results for the subID\_ncBIT and subBIT\_ncID are fully equal, which is in compliance with the expected results, as presented in Sec. \ref{sec:subnet-organization}. 
% This occurs due to the manner in which the encoding schemes together with its configuration is chosen. 
\begin{table}[]
    \centering
     \begin{tabular}{|>{\centering\arraybackslash}m{0.2\linewidth}|>{\centering\arraybackslash}m{0.2\linewidth}|>{\centering\arraybackslash}m{0.2\linewidth}|>{\centering\arraybackslash}m{0.2\linewidth}|} \hline
       Encoding Scheme  & Minimum Speedup (semi,fully) & Maximum Speedup (semi,fully) & Peak Average Speedup (semi,fully) \\ \hline
       % subID\_ncBIT & (0.7,0.5) & (55.8,30.4) & (11.8,8.1) \\ \hline
       % subBIT\_ncID & (0.7,0.5) & (55.8,30.4) & (11.8,8.1) \\ \hline
       % subBIT\_ncBIT & (0.7,0.5) & (56.2,30.0) & (9.0,5.9)\\  \hline
       subID\_ncBIT & (0.7,0.5) & (55.8,30.4) & (16.5,12.5) \\ \hline
       subBIT\_ncID & (0.7,0.5) & (55.8,30.4) & (16.5,12.5) \\ \hline
       subBIT\_ncBIT & (0.7,0.5) & (56.2,30.0) & (12.8,9.22)\\  \hline
    \end{tabular}
    \caption{Minimum, maximum, and peak average speedup values obtained for each encoding scheme using the combined designs.}
    \label{tab:combined_speedup}
\end{table}
\subsection{Combined Speedup}\label{sec:Combined_speedup}
\begin{figure*}[ht!]
    \centering
    \includegraphics[width = \textwidth]{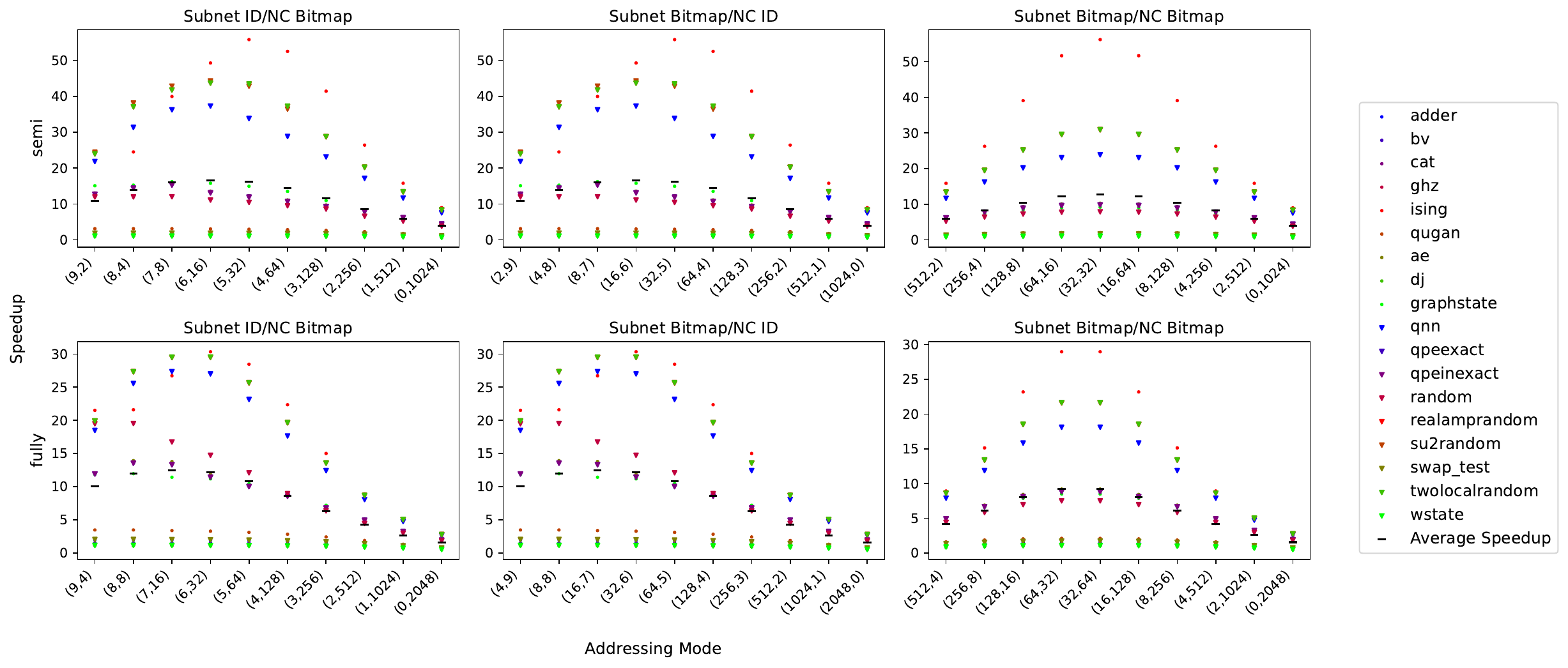}
    \caption{Speedup results achieved by the combined hardware and compiler solutions. Each tick of the x-axis corresponds to a particular addressing mode, labeled $(W_S, W_{NC})$, where $W_S$ and $W_{NC}$ are the subnet and node controller address widths, respectively. For each addressing mode, colored data points mark the speedup value for a particular benchmark, while the black data point indicates the average speedup. The top row shows speedup values in the semi-distributed mode, while the bottom row shows speedup values in the fully distributed mode.}
    \label{fig:total_speedup_combined}
    % \Description{This figure shows the speedup achieved by the combined compiler and hardware optimizations for each benchmark, as well as the average speedup across all benchmarks, for every combination of encoding scheme and distribution mode. The trends in speedup across instruction address width is the same here as it was for the hardware speedup results.}
\end{figure*}

The cumulative speedup for the compiler combined with the hardware design are presented in this section. The results for the different encoding schemes and distribution modes are presented in Fig. \ref{fig:total_speedup_combined}.
% The plots show the positioning of each possible addressing mode in $\rho$-$\delta$ space using black data points. The data points are labeled $(W_S, W_{LC})$ according to the subnet address width, $W_S$, and node controller address width, $W_{LC}$ of the given addressing mode.
The minimum, maximum, and peak average speedup results for each encoding scheme are presented in Tab. \ref{tab:combined_speedup}.

% The subID\_ncBIT encoding scheme results present a minimum speedup of 2 (2) and a maximum speedup of 90 (68) for the semi- (fully) distributed modes. The figure presents a speedup peak at (5,32) for both distribution modes. 

% The subBIT\_ncID encoding scheme results present a minimum speedup of 2 (2) and a maximum speedup of 68 (68) for the semi- (fully) distributed modes. The figure presents a speedup peak at (5,32) for both distribution modes. 

% The subBIT\_ncBIT encoding scheme results present a minimum speedup of 2 (2) and a maximum speedup of 70 (66) for the semi- (fully) distributed modes. The figure presents a speedup peak at (32,32) and (32,64)/(64,32) for the semi- and fully distributed modes respectively, with an average peak speedup of 49 (50) for the semi- (fully) distributed modes.  

These results show that, for each encoding scheme, the use of a network structure in hardware tuned to that encoding scheme together with a compiler aware of this structure is capable of achieving speedup for every benchmark. 
% Due to the fact that the speedup gained is a combination of the previous two results, it is logical that similar behaviour can be seen from these results. 
The hardware design dictates the trends for when speedup increases and decreases, while the hardware and compiler together contribute to the large absolute speedup values. Furthermore, the speedup is still dependent on the type of quantum algorithm as well as the encoding scheme and distribution mode used. We also still observe that the subID\_ncBIT and subBIT\_ncID encoding schemes produce identical results.
% The same is true for the dependency on the encoding schemes and distribution modes.

% At last, the results for the subID\_ncBIT and subBIT\_ncID again show that they are fully equal, which is in compliance with the expected results. This occurs due to the manner in which the encoding schemes together with its configuration is chosen. 
\bigskip
\section{Conclusion} \label{sec:conclusion}
% In this paper, we presented a method to reduce execution times of quantum algorithms on distributed quantum systems by creating a hardware design for parallel execution and a compiler that increases the amount of instructions that can be executed in parallel or pipeline. 
In this paper, we presented a novel hardware design and a compiler which are capable of significantly reducing the execution times of quantum programs on distributed quantum systems. Our hardware design consists of a flexible network configuration which can use different instruction address encoding schemes to support different types and amounts of parallel instruction execution. Our compiler increases the number of instructions that can be pipelined or executed in parallel by re-ordering logical instructions, decomposing them intelligently into physical instructions, and marking physical instructions that can be executed in parallel.

% Our hardware design enables parallel execution of instructions with identical parameters on separate nodes of a distributed system. The hardware design can be flexibly configured to use different schemes for encoding instruction addresses such that different types and amounts of parallel execution are supported. However, support for parallel execution in hardware introduces additional system overhead.

% We also created a compiler which is capable of increasing the amount of instructions that can be pipelined or executed in parallel. The compiler re-orders logical instructions in a quantum program to increase opportunities for parallel execution, intelligently decomposes logical instructions into physical instructions, and subsequently marks the physical instructions that can be executed in parallel based on the underlying hardware configuration. 
In our work we evaluated the compiler and hardware to work on two different configurations of logical to physical mapping. The first configuration maps the physical qubits to two nodes (semi-distributed) and the second configuration maps the qubits to 4 nodes (fully distributed). We evaluated the speedup of the compiler, hardware, and their combination using a runtime calculator and a set of benchmark quantum programs. The compiler-only solution yielded a speedup ranging from 1.04x to 13.55x, with an average of 6.11x and 3.88x for the semi- and fully distributed modes respectively. The hardware provided an additional speedup of 0.55x to 6.0x relative to the compiler, with a peak average of 1.96x and 2.48x %1.5x and 1.7x 
for the semi- and fully distributed modes respectively. Finally, the combined solution achieved a speedup between 0.7x and 56.2x, with a peak average speedup of 16.5x and 12.5x % 11.8x and 8.1x 
for the semi- and fully distributed modes respectively.  It is important to note that the lowest speedup value will only be observed if the network is partitioned poorly. Every algorithm in the benchmark can be executed with an improved speedup if the configuration is chosen correctly. Overall, it was found that each of the proposed encoding schemes was capable of speeding up every quantum program if the addressing mode corresponding to peak average speedup was selected. Finally, the amount of speedup achieved for a given program was significantly influenced by the type of program (e.g., entangled state preparation).

In the future, our research can be extended by identifying alternative encoding schemes or hardware designs which can offer higher speedup. Furthermore, it should be investigated how speedup can be improved by a compiler which is capable of reordering instructions both upwards and downwards in the program sequence. Another interesting extension is to determine if reordering can be improved by making the compiler aware of the issue and execution times for individual instructions. Despite these potential areas of improvement, the current implementation already expands the capabilities of hardware and compilers to achieve high speedup in the execution of quantum programs. Beyond yielding higher-performance computing systems, reduced execution times of quantum programs can also mitigate the impacts of noise on computations, enabling quantum programs to be executed with greater fidelity.

\section{Acknowledgements}
We gratefully acknowledge support from the joint research program “Modular quantum computers” by Fujitsu Limited and Delft University of Technology, co-funded by the Netherlands Enterprise Agency under project number PPS2007.
% \end{acks}

\clearpage 
\bibliographystyle{IEEEtran}
\bibliography{report.bib}

\end{document}